\documentclass[a4paper,11pt]{article}

\usepackage{amsfonts,amsmath,amssymb}
\usepackage{yhmath}
\usepackage{physics}
\usepackage{tensor}
\usepackage{array}
\usepackage{mathtools}
\usepackage{braket}
\usepackage{slashed}
\usepackage{cancel}
\usepackage{leftidx}
\usepackage{stackrel}
\usepackage{cite}

\newcommand{\mL}{\mathcal{L}}

\newcommand{\no}{\nonumber}
\newcommand{\der}{\partial}
\newcommand{\be}{\begin{equation}}
\newcommand{\ee}{\end{equation}}

\newcommand{\rd}{{\rm d}}
\newcommand{\myangle}{\psi}

\usepackage{enumerate}
\usepackage{bm}
\usepackage[thinlines]{easytable} 
\usepackage{tabu}
\usepackage[normalem]{ulem}

\usepackage{xcolor}
\colorlet{darkblue}{blue!70!black}
\usepackage[colorlinks=true,urlcolor=darkblue,linktocpage=true,linkcolor=darkblue,citecolor=green]{hyperref}

\usepackage{graphicx,epstopdf}
\usepackage{subcaption}

\usepackage{tikz}
\usepackage[compat=1.1.0]{tikz-feynman}

\graphicspath{{./figures/}}

\begin{document}
\title{Scattering off Chamblin-Reall Branes}
\author{Dongsheng Ge$^{\dagger,\uplus}$, and Christopher P. Herzog$^{\ddag}$}
\date{\today}

\maketitle

\vspace*{0.1cm}
\centerline{$^\dagger$\it Yau Mathematical Sciences Center,}
\centerline{\it Tsinghua University, Beijing 100084, China}
\centerline{$^\uplus$\it Department of Physics, The University of Osaka,}
\centerline{\it
Machikaneyama-Cho 1-1, Toyonaka 560-0043, Japan}
\centerline{$^\ddag$\it  Department of Mathematics, King's College London,}
\centerline{\it The Strand WC2R 2LS, England}


\vskip 1cm 


\begin{abstract}

We study the linearized scattering of dilaton-graviton waves from a thin brane in three-dimensional spacetime. Holographically, the setup models scattering from an interface in a family of strongly coupled theories related to dimensional reductions of higher-dimensional $AdS_{d+2}$ gravity. Unlike the pure $AdS_3$ case, for $d>1$ the physical bulk mode allows incident radiation to be redistributed into reflected, transmitted, and evanescent components. For the $d=2$ background, we obtain a controlled solution in which the interface acts like a rough translucent window, producing diffuse angular scattering and absorption into surface modes. From the dual perspective, the scattering process is suggestive of dissipative flow toward the infrared.
For $d=4$, the same analysis reveals a sensitivity to the infrared boundary condition, suggesting that the singular zero-temperature geometry must be regulated in order to have a well-defined scattering process. 
The structure of the equations nevertheless suggests that a regulated $d=4$ problem may exhibit the same qualitative redistribution of incident flux.

\end{abstract}

\thispagestyle{empty}
\newpage
\tableofcontents


\section{Introduction}

Scattering from interfaces is one of the simplest dynamical probes of a defect quantum field theory. In two-dimensional conformal field theories, reflection and transmission coefficients are constrained by stress-tensor two-point functions  \cite{Quella:2006de,Meineri:2019ycm}, and the same process has a holographic realization in which gravitational perturbations scatter from a thin brane in $AdS_3$  \cite{Bachas:2020yxv}. Motivated by this example, we ask how the scattering problem changes when the bulk dual is no longer pure three-dimensional gravity but includes a propagating scalar degree of freedom.

\vskip 0.1in
\noindent
Interfaces and defects are especially well understood in two-dimensional conformal field theory, where conformal symmetry and, in some examples, integrability give powerful constraints on scattering and transport
(see e.g.~\cite{kane1992transport,fendley1995exact,Oshikawa:1996dj}). Much less is known once conformal symmetry is broken, or when one asks for higher-dimensional analogues. Holography provides a tractable setting in which such questions can be reformulated as classical gravitational scattering problems. Our goal here is to construct one such problem in a simple deformation of the $AdS_3$ thin-brane setup.\footnote{%
 See \cite{Liu:2025gle,Banerjee:2025agg} for  different non-conformal deformations.
}

 \vskip 0.1in
 \noindent  
In particular, we study this question in three-dimensional Chamblin-Reall dilaton gravity \cite{Chamblin:1999ya}. The model is a  deformation of the $AdS_3$ thin-brane setup: one adds a dilaton with an exponential potential, and constructs a thin brane separating two Chamblin-Reall regions with different length scales. These backgrounds are useful because they can also be obtained by dimensional reduction of pure gravity with negative cosmological constant from $AdS_{d+2}$  \cite{Chamblin:1999ya}. Thus the same calculation admits two complementary interpretations: as scattering in a nonconformal 1+1-dimensional holographic model with hyperscaling violation \cite{Hartnoll:2016apf}, or as a reduced description of a higher-dimensional conformal problem.

 \vskip 0.1in
 \noindent  
The key difference from the pure $AdS_3$ case is the spectrum of linearized fluctuations. Pure three-dimensional gravity has no propagating gravitons; the modes scattered in the $AdS_3$ interface problem are boundary-graviton-like gauge modes whose near boundary behavior changes the boundary stress tensor. In the Chamblin-Reall model, by contrast, the linearized spectrum contains a physical mixed dilaton-graviton wave. Scattering this mode from the brane produces a richer optical problem, with reflected and transmitted radiation, angular redistribution, and evanescent surface modes.\footnote{%
Few works in general relativity refer explicitly to evanescent gravitational waves.
Ref.\  \cite{Golat:2019aap} explores their effect on test masses and their relevance for subwavelength gravitational wave sources.  
The graviton bound state on a Randall-Sundrum brane \cite{Randall:1999vf,Seahra:2005wk} is arguably another example of an evanescent gravitational wave
although this terminology is not standard in that context.
}

\vskip 0.1in
\noindent
Our main result, presented in section \ref{sec:IJ}, is an analysis of the scattering problem for the $d=2$ and $d=4$ Chamblin-Reall backgrounds (which have uplifts to $AdS_4$ and $AdS_6$ respectively).  
For $d=2$, the scattering problem admits a clean optical interpretation: an incident dilaton-graviton wave is redistributed into reflected, transmitted, and evanescent components. Rather than producing only specular reflection and direct transmission, the brane emits reflected and transmitted radiation over a range of angles that we determine explicitly.
From the dual perspective, the scattering transfers support from modes localized near a fixed RG scale into modes propagating toward the infrared, suggesting a form of dissipative relaxation.
For $d=4$, the same formalism exposes a sensitivity to the IR boundary condition, suggesting that the zero-temperature Chamblin-Reall singularity must be regulated before the scattering problem is fully well defined. The structure of the equations nevertheless suggests that a regulated $d=4$ problem should exhibit the same qualitative redistribution of incident flux.

\vskip 0.1in
\noindent
There are two important limitations to our analysis.  The Chamblin-Reall spacetimes have curvature singularities.  While these singularities are of a ``good kind'' and can be screened by black hole event horizons (or dually by introducing a nonzero temperature to the field theory) \cite{Gubser:2000nd}, our analysis in this work is at zero temperature.  Despite the presence of these singularities, there has been a great deal written about these geometries in the applied AdS/CFT community over the years (for a sampling of the literature, see e.g.\ \cite{Gursoy:2007cb,Gursoy:2007er,Gubser:2008sz,Charmousis:2010zz,Gursoy:2015nza,Betzios:2017dol,Betzios:2018kwn}).
Next, because the calculation is performed in a three-dimensional reduced geometry, the scattering angle is an angle in the $\rho x$-plane rather than an incidence angle along independent field-theory spatial directions. A higher-dimensional uplift would be needed to study more general field-theory kinematics.

\vskip 0.1in
\noindent
An outline of the paper is as follows.  In section \ref{sec:CR}, we review the Chamblin-Reall geometries in three-dimensional dilaton gravity and their uplifts to $AdS_{d+2}$.
We further construct thin branes that separate two such geometries characterized by different length scales $\ell$.  
In section \ref{sec:linearized}, fixing a radial gauge, we completely characterize the linearized fluctuations about these geometries.  There is a dilaton-graviton wave in addition to three residual gauge transformations that preserve the radial gauge choice.  We also write down an effective action for the dilaton-graviton waves and from it construct a conserved flux.  Finally in section \ref{sec:IJ}, we solve the Israel junction condition and other continuity conditions across the thin brane in order to characterize the scattering of dilaton-graviton waves off of the thin brane at linearized order.   Appendices provide some supporting details for the calculation in the main body of the paper.   Appendix \ref{sec:EA} provides a one-variable effective action for the brane angle.  This effective action allows us to deduce a stability condition that is useful for cutting down the physical parameter space of our scattering process.   
Appendix \ref{sec:linearization} provides a formalized way of carrying out linearized perturbation theory on the Israel junction condition.   Appendix \ref{sec:solns} records the embedding-function solutions induced by the dilaton-graviton waves.  
Appendix \ref{sec:cubic} lists the roots of a cubic equation important in characterizing solutions for the $d=4$ case.

\section{Review of CR geometry}
\label{sec:CR}

Domain walls in the geometries we consider were studied in detail by Chamblin and Reall 
\cite{Chamblin:1999ya} over two decades ago.  
There are, however, important differences in aims and approach.
While their solutions exist for general spacetime dimension $D$, here we fix
$D=3$ to keep the scattering calculation to come manageable.  
Their interest was in modeling time-dependent cosmologies through placing moving domain walls in static backgrounds.
They further assume a reflection symmetry of the bulk spacetime about the domain wall.
Our setup instead generalizes the static domain walls of ref.\ \cite{Bachas:2020yxv} to theories with an additional dilaton.
These static domain walls glue two different spacetimes together and are the targets from which the dilaton-graviton
waves of section \ref{sec:linearized} scatter.

\vskip 0.1in
\noindent
Before getting to the domain walls, we spend some time reviewing the spacetime in the absence of a domain wall but in the presence of a black-brane horizon, which introduces a nonzero temperature to the dual field theory.  
Although the scattering calculation below uses the zero-temperature geometry, the  nonzero temperature solution is useful for two reasons.
First, it explicates the parameter $d$, connecting the range of $d$ to the range of black brane solutions with positive specific heat and also to the uplifted
$AdS_{d+2}$ spacetimes.  Second, it allows us to argue that the singularities in these geometries are of a ``good kind'' that can be shielded by a horizon.

\vskip 0.1in
\noindent
The starting point for both us and ref.\ \cite{Chamblin:1999ya} is a three-dimensional Einstein-Hilbert plus dilaton action with an exponential potential for the dilaton:
\begin{equation}
S_{\rm bulk} = \frac{1}{2 \kappa^2} \int \rd^3x  \sqrt{-g} \left[ R - \frac{1}{2} (\partial \phi)^2 - \frac{1}{l^2} V_0 e^{\gamma \phi} \right] \ .
\end{equation}
We will see below the thermodynamically stable range is $0 < \gamma < \sqrt{2}$ with $\gamma=0$ understood as the $AdS_3$
limit and the endpoint $\gamma = \sqrt{2}$ corresponding to taking $d \to \infty$.   
The quantity $l$ sets a length scale, and $\kappa$ sets the gravitational coupling strength.

\vskip 0.1in
\noindent
While we ultimately will use geometries without a black-brane horizon, to get a better sense
of the physics governing these systems, we start by looking for solutions with a horizon:
\begin{equation}
\rd s^2 = e^{2 A(r)} (- f(r) \rd t^2 + \rd x^2 ) + e^{2 B(r)} \frac{l^2 \rd r^2}{ f(r)} \ .
\end{equation}
The horizon by construction sits at $r=r_0$ such that $f(r_0) = 0$. 
In our conventions, $r$ and $\phi$ are dimensionless but $t$ and $x$ have dimensions of length.
Making the gauge choice $\phi(r) = r$, there are solutions of the form
\begin{eqnarray}
&& A(r) = -\frac{r}{\gamma} \ , \; \; \; B(r) = - \frac{\gamma r}{2} \ ,  \; \; \; V_0 = \frac{1}{2} - \frac{2}{\gamma^2} \ , \\
&& f(r) = 1 - e^{\left(\frac{2}{\gamma} - \frac{\gamma}{2} \right) (r-r_0) }\ .
\end{eqnarray}
Provided $0 < \gamma < 2$, the horizon can be removed by taking $r_0 \to \infty$.  This behavior of the horizon 
suggests $r \to \infty$ is the interior of the geometry and
$r \to -\infty$ is the boundary.  In the special case $\gamma = \sqrt{2}$, the warp factors $A$ and $B$ become identical.  
Without the black-brane horizon, there is a naked curvature singularity in the Ricci scalar, $R \sim e^{r \gamma}$, in the limit $r \to \infty$.

\vskip 0.1in
\noindent

Let us explore the thermodynamics of this black brane solution in more detail.
Using the standard trick that in the Wick rotated version of this spacetime $t = - i \tau$, the horizon $r=r_0$ should be a smooth point in the geometry,
periodicity in Euclidean time $\tau$ reveals the Hawking temperature to be
\begin{equation}
T = \frac{1}{\beta} 
=\frac{ (4-\gamma^2)}{8 \pi \gamma  l} e^{-r_0 \left(\frac{1}{ \gamma} - \frac{\gamma}{2}\right)}  \ .
\end{equation} 
The entropy should go as the horizon area, and so the entropy density scales as 
\begin{equation}
s \sim e^{A(r_0)}  =e^{-r_0/\gamma}  \sim T^{\frac{2}{2 - \gamma^2} } \ .
\end{equation}
Positive specific heat of the geometry thus requires $\gamma < \sqrt{2}$, which is the thermodynamically stable range we will use below.
That the presence of any slightly nonzero temperature introduces a horizon which shields the singularity makes these singularities ``good'' in the sense of ref.\ \cite{Gubser:2000nd}.  

\vskip 0.1in
\noindent
 In the language of AdS/CFT, we can attach these thermodynamic quantities to a dual field theory living at the boundary of the spacetime.  Or alternatively, we could try to work on a stretched horizon close to the actual event horizon.  In either case, compactifying the $x$ coordinate, there is an associated spatial volume $V$ (really just a length) we can associate with the geometry, and a total entropy $S = sV$.   Using the thermodynamic relation between the entropy and the Helmholtz free energy $S = -\frac{\partial F}{\partial T}$, $F$ should scale as 
\begin{equation}
F  = -\frac{2-\gamma^2}{4 - \gamma^2} S T \sim T^{\frac{4-\gamma^2}{2-\gamma^2}} \ .
\end{equation}    
Note the free energy is directly a product of the pressure and the volume $F = - p V$ in the thermodynamic limit where we work.
By definition $\epsilon + p = s T$.  Thus the energy density is $\epsilon = \frac{2}{4 - \gamma^2} s T$.  The speed of sound squared is 
\begin{equation}
c_s^2 = \frac{\partial p}{\partial \epsilon} = 1 - \frac{\gamma^2}{2} \ ,
\end{equation}
 which is greater than zero given the bound $\gamma < \sqrt{2}$.
 
 \vskip 0.1in
 \noindent
  There is a strong connection between these Chamblin-Reall spacetimes and $AdS_{d+2}$ geometries.
  As noted by Chamblin and Reall  \cite{Chamblin:1999ya}, the dilaton geometries can be obtained by compactifying
  a pure $(d+2)$-dimensional  gravity action with negative cosmological constant along a $(d-1)$-dimensional torus.
We can deduce the relation between $d$ and our parameters in the following way.
 From an AdS/CFT standpoint, a black brane geometry in $AdS_{d+2}$ should have a dual description as a thermal $(d+1)$-dimensional conformal field theory.  In that context, the speed of sound is restricted by the tracelessness of the stress tensor to always be $c_s^2 = \frac{1}{d}$.  
 (We distinguish the dimension of the spacetime geometry $D=3$ from the parameter $d$ which characterizes the dimension of the higher-dimensional conformal field theory before torus reduction.)

 \vskip 0.1in
 \noindent
 In what 
 follows, it is convenient to parametrize our solutions by $d$ instead of $\gamma$ or $c_s$:
 \begin{equation}
 \gamma = \sqrt{2} \sqrt{1 - c_s^2} \ , \; \; \; c_s^2 = \frac{1}{d} \ .
 \end{equation}
 There are two special limits.  The first is $d \to 1$ in which we recover pure three-dimensional gravity with no dilaton and a negative cosmological constant.
 Another special limit is $d \to \infty$, where we recover the upper bound $\gamma = \sqrt{2}$.  In this second case, the zero temperature 
 metric becomes conformally flat in the gauge where the dilaton is linear in $r$.  
 The $d=1$ case was the original inspiration for this study \cite{Bachas:2020yxv}.  We do not consider the $d \to \infty$ limit further here
 \cite{Betzios:2018kwn,Emparan:2020inr}.
In later sections we focus on $d=2$ and $d=4$, where there should be a relation between the results we obtain and uplifted solutions for $AdS_4$ and $AdS_6$ which in turn should have dual descriptions in terms of three-dimensional or five-dimensional conformal field theories.
The reason for the specialization to $d=2$ and 4 is a special simplification that happens for fluctuations around these two spacetimes, as we will see later.  

\vskip 0.1in
\noindent
Having introduced the black brane through the function $f(r)$ in order to understand the physics of our setup more deeply, we now discard it.  
We will work with a solution with $f(r) = 1$, safe in the knowledge that the singularity at $r\to \infty$ can be screened with a small temperature
and that there is a sense in which what we find should be related to similar problems in $AdS_{d+2}$ geometries.
In what comes next, we design an interface geometry and study fluctuations inside of it.  The interface will be simplest to construct using coordinates that make manifest the geometry is conformally flat.  Also, as the fluctuations travel along lightlike geodesics, their description will be simpler with such coordinates.

\vskip 0.1in
\noindent
To make things conformally flat, we must give up the gauge where $\phi = r$: 
\begin{equation}
\rd\rho = e^{B-A} l \rd r \; \; \implies \; \; \rho =
l \gamma d \exp \left( \frac{r}{\gamma d}  \right)\ . 
\end{equation}
In this coordinate system,
the conformally flat line element and dilaton are
\begin{eqnarray}
\label{cfmetric}
\rd s^2 &=& \left( \frac{\gamma l d}{ \rho} \right)^{2d} (-\rd t^2 + \rd x^2 + \rd \rho^2) \ , \\
\label{bgdilaton}
\phi(\rho) &=& \gamma d \log \left( \frac{\rho}{l \gamma d} \right) \ .
\end{eqnarray}
Unfortunately, the solution in this form makes the $d\to 1$ limit obscure.  To recover it, one must simultaneously send $l \to \infty$ and $\gamma \to 0$ while keeping $\gamma l$ fixed.  With this procedure implemented, the dilaton vanishes, and we recover $AdS_3$ in the Poincar\'e patch, as expected. 
To make the $d \to 1$ limit more straightforward, and also to make the notation more compact, we redefine the length scale at this point
\begin{equation}
\label{Lredef}
\ell \equiv l \gamma d \  .
\end{equation}
We use $l$ only in the intermediate black-brane parametrization and use $\ell$ throughout the rest of this work.
 With this definition, the conformal factor remains finite in the $d\to 1$ limit.

\subsection{The Interface}

Our interfaces sit at a fixed angle in the $\rho x$-plane and are  designed to separate metrics with different values of $\ell$, $\ell_R \neq \ell_L$. 
To construct these interfaces, we add 
 a thin brane with the two-dimensional action
\begin{equation}
S_{\rm brane} = \frac{1}{2\kappa^2} \int_I \rd^2 x \sqrt{-h} \left( 2 (K_R - K_L) + \mu e^{\alpha \phi } \right) \ ,
\end{equation}
where $\mu$ is a scale governing the strength of the dilaton-dependent brane tension 
and $\alpha$ is a dimensionless number fixing its growth with $\phi$.
$K_R$ and $K_L$ are the traces of the extrinsic curvature on either side.  
The induced metric on the brane is $h_{ab}$.  
To make the notation more compact, we use a bracket to indicate the difference between the right and left sides of the interface in what follows:
\begin{equation}
[X] \equiv X_R - X_L \ .
\end{equation}

\vskip 0.1in
\noindent
There are several matching conditions.  The first two are continuity of the metric and the dilaton, $\phi_L = \phi_R$ and $(h_L)_{ab} = (h_R)_{ab}$, at the interface.
More precisely, continuity of the metric means that the brane metrics $(h_{R/L})_{ab}$ induced from the bulk metrics on the right and left hand sides
agree.  
The second two matching conditions are imposed by having a good variational principle for the classical action.
On the metric, we have the Israel junction condition, and on the dilaton, a jump condition on the normal derivative.
Assembled, the four matching conditions, written with the bracket notation, are
\begin{align}
{} [ h_{ab}] &= 0 \ , \; \; \; & [\phi] &= 0 \ , \\
{} [ K_{ab} - h_{ab} K ] &=  \frac{\mu}{2} e^{\alpha \phi} h_{ab}  \ , \; \; \; & [ \partial_n \phi ] &= \mu \alpha e^{\alpha \phi} \ .
\label{Israel}
\end{align}
The unit normals $n_{L/R}^\mu$ are defined with a sign convention to be compatible with these difference relations.

\begin{figure}
\begin{center}
 \includegraphics[width=3in]{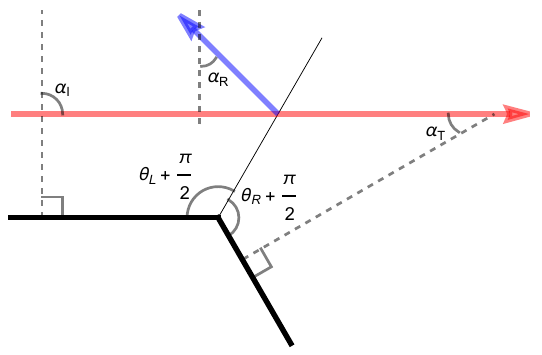}
\end{center}
\caption{
 Basic setup with angles identified.
}
\label{fig:basicsetup}
\end{figure}

\vskip 0.1in
\noindent
Our interface sits at a constant angle in the $x \rho$-plane, beginning at $(x,\rho) = (0,0)$ and stretching into the positive $\rho$ direction (see figure \ref{fig:basicsetup}). 
We pass from the $(t,x,\rho)$ coordinate system to the polar coordinates $\sigma^2 = \rho^2 + x^2$ and $\tan \theta_L = \frac{x}{\rho} = - \tan \theta_R$.
The conventions, where the angle measures the deviation of the interface from the $\rho$ direction, are consistent with ref.\ \cite{Bachas:2020yxv}.  Positive $\theta_R$ and $\theta_L$ correspond to opening angles greater than ninety degrees.

\vskip 0.1in
\noindent
Continuity already strongly constrains the interface geometry.
The dilaton and metric depend only on the radial coordinate  through the combination $\rho/\ell$. Along the brane in polar coordinates,
we are led to compare $\rho / \ell = ( \sigma  / \ell) \cos \theta$.  
 The dilaton for example becomes 
\begin{equation}
\phi = \gamma d \log \left( \frac{ \sigma \cos \theta}{\ell}   \right) \ .
\end{equation}
Continuity of the dilaton (and also the metric) then implies
\begin{equation}
 \frac{\cos \theta_L}{\ell_L} =  \frac{ \cos\theta_R }{\ell_R} \equiv \frac{1}{\ell_B}\ \ ,
\end{equation}
where we have introduced a common brane length scale $\ell_B$. 
To satisfy the remainder of the conditions -- Israel junction and dilaton jump -- we tune the dilaton-dependent brane tension such that
\begin{eqnarray}
\label{alphamurels}
\alpha 
&=& \frac{\gamma}{2} \ ,  \; \; \;
\mu = \frac{2d}{\ell_B}   \left( \tan \theta_R + \tan \theta_L \right)  \ .
\end{eqnarray}
The value of $\alpha$ is fixed by requiring that the equations scale homogeneously as a power of $\sigma$.\footnote{%
 The choice of $\alpha$ also guarantees the uplifted brane in $AdS_{d+2}$ has constant tension.
}
Positive tension requires $m \equiv \tan \theta_R + \tan \theta_L>0$. 
In appendix \ref{sec:EA}, we describe an additional stability criterion, which further restricts the useful region of parameter space.

\vskip 0.1in
\noindent
It is useful to record one consistency check on the matching problem. The metric continuity and Israel equations appear to give $D(D-1)$ conditions, and the dilaton adds two more. However, not all of these conditions are independent. The tangential divergence of the Israel equation is fixed by the mixed normal-tangential Einstein equation $G_{na}$ 
and gives the momentum-balance equation for the interface stress tensor. In the present scalar system this balance equation follows from dilaton continuity together with the jump condition for $\partial_n \phi$. Thus the junction conditions should not be viewed as an arbitrary overdetermined set of equations; their tangential divergence is constrained by the bulk equations of motion.

\vskip 0.1in
\noindent
To see why the divergence of the Israel condition is a constraint on the conditions, recall that in an ADM decomposition of the metric the $G_{an}$ portion of the Einstein equations can be written in the form
\begin{eqnarray}
\nabla^b [ K_{ba} - h_{ba} K ]  &=& [ T_{na} ] \ .
\end{eqnarray}
The Israel condition on the other hand implies
\begin{equation}
\nabla^b [ K_{ba} - h_{ba} K ] =  \nabla^b \frac{\mu}{2} e^{\alpha \phi} h_{ba} \ ,
\end{equation}
which in turn tells us that
\begin{equation}
[ T_{na} ] =  \nabla^b \frac{\mu}{2} e^{\alpha \phi} h_{ba} \ .
\end{equation}
This $\frac{\mu}{2} e^{\alpha \phi} h_{ab}$ is really the interface stress tensor.
This last condition is nothing but momentum conservation: the divergence of the interface stress tensor is accounted for by momentum leakage into the bulk.
In this particular case, one can check the last statement by inserting the specific bulk form for the stress tensor $T_{na} \sim (\partial_n \phi )(\partial_a \phi)$ and observing that this momentum balance constraint follows from dilaton continuity and the dilaton jump condition.

\section{Dilaton-graviton waves in CR geometry}
\label{sec:linearized}

The result of this section is that radial gauge leaves three residual gauge modes and one physical propagating mode. The residual gauge modes are needed later to satisfy the interface matching conditions near $\rho = 0$, while the physical mode is the dilaton-graviton wave whose flux is scattered by the brane.
More precisely, our radial gauge choice sets $\delta g_{\rho \mu} = 0$, and the three residual gauge transformations preserve this condition. 
These four classes of solution can be viewed as small perturbations around the conformally flat metric (\ref{cfmetric})
 and dilaton solution (\ref{bgdilaton}) found above.

%
%

\vskip 0.1in
\noindent
A direct approach is to expand Einstein's equations and the equation of motion for the dilaton to first order in the fluctuations and solve them.  To understand the physical meaning of the four resulting classes of modes, however, it is more illuminating to investigate the residual gauge solutions directly and then to establish, via a quadratic action,  a physical flux associated with the dilaton-graviton waves.

\subsection{Solving the equations of motion}
\label{sec:solving}

Fixing radial gauge $\delta g_{\mu \rho} = 0$ and given translational invariance in the $x$ and $t$ direction, we look for plane wave solutions of the form
$\delta\phi = e^{-i \omega t + i k x} \varphi(\rho)$, $\delta g_{tt} = (g_+(\rho) + g_-(\rho))e^{-i \omega t + i k x}$,
$\delta g_{xx} = (g_+(\rho) - g_-(\rho))e^{-i \omega t + i k x}$, and $\delta g_{xt} =g_2(\rho) e^{-i \omega t + i k x}$.
In this notation, $g_-$ is the trace degree of freedom of the metric.  
For those wishing to skip ahead, a quick summary is given by the solutions (\ref{g2}), (\ref{gplus}), (\ref{varphi}), and (\ref{gminus}) for the fluctuations $g_2$, $g_+$, $\varphi$, and $g_-$ respectively.  The constants $c_2$, $c_+$, and $c_1$ multiply residual gauge transformations while $c$ multiplies the physical  dilaton-graviton wave.

\vskip 0.1in
\noindent
In more detail, the linearized equations of motion about the background have the form
\begin{eqnarray}
{\rm EOM}_\varphi &=& \rho^d \left( \frac{\varphi'}{\rho^d} \right)' - \frac{\gamma d}{ \rho} \left( \left(\frac{\rho}{\ell} \right)^{2d} g_- \right)' + \left( \omega^2 - k^2 + \frac{2(d^2-1)}{\rho^2} \right) \varphi \ , \\
{\rm EOM}_{tt} &=& \frac{i}{\omega \rho^d} ( \rho^d {\rm EOM}_{t \rho} )'  - \frac{k}{\omega} {\rm EOM}_{tx} \\
{\rm EOM}_{tx} &=& \frac{1}{2 \rho^2} \left(\rho^2 g_2'' + 3d \rho g_2' + \gamma^2 d^2 g_2 \right) \ , \\ 
{\rm EOM}_{t \rho} &=& \frac{i \omega}{2} \biggl(  \frac{1}{\rho^{2d}} \left(\rho^{2d} \left( g_- -  g_+   - \frac{ k}{\omega} g_2\right)\right)' - \frac{\gamma d}{\rho} \left( \frac{\ell}{\rho} \right)^{2d}  \varphi \biggr)  \ , \\
{\rm EOM}_{xx} &=& -\frac{i}{\omega \rho^d} \left( \rho^d {\rm EOM}_{t\rho} \right)' + \frac{k}{\omega} {\rm EOM}_{tx}
+ \frac{1}{\rho^2} \left( \rho^2 g_+'' + 3d \rho g_+'  + \gamma^2 d^2  g_+ \right) \  , \\
{\rm EOM}_{x \rho} &=& -\frac{i k}{2} \biggl(  \frac{1}{\rho^{2d}} \left(\rho^{2d} \left( g_- + g_+   + \frac{ \omega}{k} g_2\right)\right)' - \frac{\gamma d}{\rho} \left( \frac{\ell}{\rho} \right)^{2d} \varphi \biggr)  \ , \\
{\rm EOM}_{\rho \rho} &=& -\frac{d}{\rho} g_-' + \frac{\gamma d}{2} \left( \frac{\ell^{2d}}{\rho^{2d+1}} \right)  \varphi' 
+ \frac{1}{2} \left( \omega^2 - k^2 - \frac{ 4d^2}{\rho^2} \right) g_- + \frac{\gamma d(d+1)}{2} \frac{ \ell^{2d}}{\rho^{2d+2}}  \varphi
\nonumber \\
&&  - \omega k g_2 - \frac{1}{2}(\omega^2 +k^2) g_+ \ .
\end{eqnarray}
Each of these expressions must vanish, and $'$ indicates $\partial_\rho$.  
The $tx$ equation fixes the off-diagonal metric component.  A particular linear combination of the $t \rho$ and $x\rho$ equations then fixes $g_+$.  The remaining coupled scalar-trace sector gives the physical dilaton-graviton waves.

\vskip0.1in
\noindent
The ${\rm EOM}_{tx}$ equation can immediately be solved to give
\begin{equation}
\label{g2}
g_2 = c_2 (\omega^2-k^2) \rho^{-2d} +  2 \omega k c_1 \rho^{-d+1} \ .
\end{equation}
where $c_1$ and $c_2$ are integration constants that we will see in the next subsection control the
amplitude of 
residual gauge transformations.
We have included $\omega$- and $k$-dependent factors  to simplify subsequent expressions.
Note that
\begin{equation}
{\rm EOM}_{\parallel} \equiv k {\rm EOM}_{t \rho} + \omega {\rm EOM}_{x \rho} = \frac{1}{i \rho^{2d}}
 \left( \rho^{2d} 
 \left( k \omega  g_+ + \frac{1}{2}(\omega^2+k^2) g_2 \right) \right)'
\end{equation}
which immediately allows us to solve for $g_+$:
\begin{equation}
\label{gplus}
g_+ = - c_1 (\omega^2 + k^2) \rho^{-d+1} + c_+ (\omega^2-k^2) \rho^{-2d} \ ,
\end{equation}
and $c_+$ is another integration constant, which controls the amplitude of the third residual gauge transformation.
Note this solution for $g_+$ is consistent with the ${\rm EOM}_{xx}$ equation of motion, 
\begin{equation}
\widetilde{\rm EOM}_{xx} \equiv  \frac{1}{\rho^2} \left( \rho^2 g_+'' + 3d \rho g_+'  + \gamma^2 d^2  g_+ \right) \ .
\end{equation}
The three constants $c_i$ parametrize the residual gauge degrees
of freedom that preserve the condition $\delta g_{\mu \rho} = 0$, as we will make clear below.

\vskip 0.1in
\noindent
Before we proceed, note that the dilaton equation of motion can be expressed in terms of the Einstein equations:
\begin{eqnarray}
\gamma {\rm EOM}_\varphi &=& \frac{2\rho}{d \ell^{2d}} ( \rho^{2d} {\rm EOM}_{\rho \rho} )' - \frac{4 i}{\omega \ell^{2d}}
\rho^{d } \left( \rho^{d }  {\rm EOM}_{t\rho} \right)'
+ \frac{2i \rho^{2d+1}}{\omega d \ell^{2d}} (\omega^2-k^2) {\rm EOM}_{t \rho}
\nonumber \\
&& + \frac{4k}{\omega} \frac{\rho^{2d}}{\ell^{2d}} {\rm EOM}_{tx} + 2\frac{\rho^{2d}}{\ell^{2d}} \widetilde {\rm EOM}_{xx} +
 \frac{2ik}{\omega} \frac{\rho^{2d+1}}{d \ell^{2d} } {\rm EOM}_{\parallel} \ .
\end{eqnarray}

\vskip 0.1in
\noindent
The physical fluctuations are encoded in the solutions for $g_-$ and $\varphi$.
Given the analysis so far, to solve for $g_-$ and $ \varphi$, it suffices to consider only ${\rm EOM}_{t\rho}$ and ${\rm EOM}_{\rho \rho}$.  These two first order equations can be combined to give either a second order equation for $g_-$ or a second order equation for $\varphi$. 
When the dust settles, we find the Hankel functions:
\begin{eqnarray}
\label{varphi}
\varphi &=& c  \sqrt{\frac{\pi}{2}} (\omega^2 - k^2)^{\frac{1}{4}}\rho^{\frac{d+1}{2}}  H^{(1)}_{\frac{d-3}{2}} (\rho \sqrt{\omega^2-k^2})  + c.c. \ ,
\\
\label{gminus}
g_- &=&  c \sqrt{\frac{\pi}{2}}    \gamma d (\omega^2-k^2)^{-\frac{1}{4}} \left( \frac{\ell}{\rho} \right)^{2d}  \rho^{\frac{d-1}{2}} H^{(1)}_{\frac{d-1}{2}} (\rho \sqrt{\omega^2-k^2}) + c.c. \ ,
\end{eqnarray}
setting the $c_1$, $c_2$ and $c_+$ to zero.  Here $c.c.$ means complex conjugation and $c$ is an integration constant.
The fluctuations for $d \in 2 {\mathbb N}$ are particularly simple because in this case the Hankel functions reduce to plane waves $e^{i \rho \sqrt{\omega^2-k^2} }$ times polynomials in $\rho$.  Among these even cases, $d=2$ and $d=4$ are simpler still because $H_{\pm \frac{1}{2}}^{(1)} (x) \sim x^{-\frac{1}{2}} e^{i x}$.  In any case, for large $\rho$, because of the asymptotic form of the Hankel functions, these wave solutions reduce, up to a redshift factor, to a plane wave traveling along a lightlike geodesic in the geometry.  
For real $\omega$, when $k^2 > \omega^2$, 
 the radial wave number becomes imaginary and the solution is evanescent in $\rho$. We will encounter complex continuations of the angular variable that realize these evanescent branches in section \ref{sec:IJ}.

\vskip 0.1in
\noindent
 If we include the gauge solutions, they modify these waves with the additions
\begin{eqnarray}
\label{varphifull}
 \varphi &\to&  \varphi - c_1 \gamma d (d+1) \ell^{-2d} \rho^{d-1} \ , \\
 \label{gminusfull}
g_- &\to& g_- +2 c_2 \omega k \rho^{-2d} + c_+ (\omega^2+k^2) \rho^{-2d}  - c_1 \rho^{-d-1} (2d(d+1) + \rho^2(\omega^2-k^2)) \ .
\end{eqnarray}

\vskip 0.1in
\noindent
There have been attempts in the literature to build a holographic dictionary around the $\rho \to 0$ limit of these fluctuations (see e.g.\
\cite{Gursoy:2015nza,Betzios:2018kwn}).  As the metric scales as $\rho^{-2d}$, we associate this behavior in the metric fluctuations as a source term in the dual field theory which alters the background metric.  The dilaton-graviton waves and the $c_1$-type gauge fluctuations have a $\rho^{-d+1}$ behavior as well.  Following the usual holographic interpretation,  this falloff has the natural interpretation as an expectation value for the dual stress tensor, which should have scaling dimension  $d+1$ in the dual field theory, at least after uplifting from two to $d+1$ spacetime dimensions. 
 More precisely, to properly identify    the  coefficient of the $\rho^{-d+1}$ falloff, one must tune the $c_1$ residual gauge transformation to zero out the leading $\rho^{-d-1}$ term in $g_-$.  This tuning will in general modify the subleading $\rho^{-d+1}$ dependence.  (In fact this step is critical in order to get an off-diagonal component of the stress tensor in this dynamical system.)  As our holographic interpretation is for the most part qualitative, we will be content to set $c_1 = 0$.

\subsection{Residual gauge transformations}

Here we establish that the integration constants $c_2$, $c_+$, and $c_1$ introduced above control the amplitude of the residual gauge transformations that preserve $\delta g_{\mu \rho} = 0$. 
Consider the infinitesimal coordinate transformation:
\begin{equation}
x^\mu \to x^\mu + \sum_{i=1}^3 a_i \xi_{(i)}^\mu
\end{equation}
where $a_i$ are numbers and 
 $\xi_{(i)}^\mu = v_{(i)}^\mu e^{-i \omega t + i k x}$.  If the coordinate transformation is a linear combination of the following three vectors,
the radial gauge condition $\delta g_{\rho \mu} = 0$ is preserved:
\begin{eqnarray}
v_{(1)}^\mu &=& \left\{1 ,0 ,0 \right\} \ , \; \; \; 
v_{(2)}^\mu = \left\{ 0, 1, 0 \right\} \ , \\
v_{(3)}^\mu &=& \left\{ \frac{i \omega \rho^{d+1}}{ \rho_0^{d}} ,   \frac{i k \rho^{d+1}}{ \rho_0^{d}}, - \frac{(d+1) \rho^d}{\rho_0^d} \right\} \ .
\end{eqnarray}
The constant $\rho_0$, which can be absorbed into the normalization of $a_3$,  is an arbitrary reference scale inserted to make the components dimensionally homogeneous.

\vskip 0.1in
\noindent
In the $d=1$ case, one of these gauge transformations played the role of the wave scattered by the interface.  
It was  the ``relativistic'' combination
\begin{equation}
\omega v_{(1)} - k v_{(2)} + \frac{i \rho_0 \omega^2}{2} v_{(3)} \ ,
\end{equation}
where further $\omega = k$.
The metric fluctuations take the simple form $\delta g_{tt} = - \delta g_{tx} = \delta g_{xx} \sim  \ell^2 \omega^4 e^{-i \omega t + i k x}$ with no $\rho$ dependence.  
Despite their origin as a coordinate transformation, the metric  falloff at the boundary is order one, $\delta g_{ab} \sim O(1)$, in a small $\rho$ expansion, and thus it contributes to the stress tensor in the dual field theory description of the system.  

\vskip 0.1in
\noindent
In our case $d>1$, the nature of this special gauge transformation at the boundary is completely different.  It will have incommensurate falloffs of order $O(\rho^{-2d})$, $O(\rho^{-d-1})$, and $O(\rho^{-d+1})$.  In the $d=1$ case, the $O(\rho^{-2d})$ and  $O(\rho^{-d-1})$ are tuned to cancel each other, 
which cannot be done more generally.  
These gauge modes alone do not appear to furnish a consistent scattering problem.
Consistent with the matching analysis of section \ref{sec:CR}, the scalar matching condition is an essential part of the junction problem, and the physical dilaton-graviton mode supplies the needed dynamical degree of freedom. 

\vskip 0.1in
\noindent
We will only have use for one particular gauge transformation in what follows, the combination
\begin{equation}
\omega v_{(1)} - k v_{(2)}
\end{equation}
which produces a purely diagonal metric fluctuation of the form
\begin{equation}
k^2 \delta g_{tt} = - \omega^2 \delta g_{xx} = \omega^2 k^2 \rho^{-2d} e^{-i \omega t + i k x} \ ,
\end{equation}
leaves the dilaton invariant, and corresponds to the solutions associated with the integration constant $c_+$ in the previous subsection.
This coordinate transformation will be useful to make the brane embedding functions continuous at $\rho = 0$.
The coordinate transformation $k v_{(1)} - \omega v_{(2)}$ on the other hand corresponds to the integration constant $c_2$ above while the slower-falloff gauge solution associated with $c_1$ corresponds to the coordinate transformation $v_{(3)}$.

\vskip 0.1in
\noindent
The three vector fields above exhaust the residual diffeomorphisms preserving radial gauge. Comparing their induced metric and dilaton variations with the general solution of section \ref{sec:solving} shows that they account precisely for the constants $c_1$, $c_2$ and $c_+$.  The remaining integration constant $c$ therefore cannot be removed by a residual gauge transformation. It labels the physical dilaton-graviton wave.

\subsection{Flux carried by the dilaton-graviton waves}
\label{sec:flux}

To identify the flux carried by the physical dilaton-graviton waves, we restrict the quadratic action to the scalar-trace sector: 
 $\delta \phi$ and the trace component $\delta g \equiv \delta g_{tt} = - \delta g_{xx}$ of the metric.  
 Although the metric and dilaton perturbations are real, it is convenient to work with complex mode functions; the physical perturbation is obtained by taking the real part. The current below should therefore be understood not as a Noether current of the full nonlinear real theory, but as the conserved Wronskian, or symplectic product, of two linearized solutions. Equivalently, it is the bilinear current obtained by pairing a mode with its complex conjugate.
 
\vskip 0.1in
\noindent
 The Lagrangian density expanded out to second order in these fluctuations takes the form
\begin{eqnarray}
{\mathcal L} = {\mathcal L}_0 + {\mathcal L}_1 + {\mathcal L}_2  + \ldots \ ,
\end{eqnarray}
where at zeroth and first order we find
\begin{eqnarray}
{\mathcal L}_0 &=&- \frac{1}{2 \kappa^2} \frac{2d (d+1) \ell^d}{\rho^{d+2}} \ , \\
{\mathcal L}_1 &=& \frac{1}{2 \kappa^2} \biggl(
2 \, \partial_\rho \left( \left(\frac{\rho}{\ell} \right)^d \partial_\rho \delta g \right) + 2 d \, \partial_\rho \left( \left(\frac{\rho}{\ell}\right)^d \frac{\delta g}{\rho} \right) \  \nonumber \\
&& - \gamma d \, \partial_\rho \left( \left( \frac{\ell}{\rho}\right)^d \frac{\delta \phi}{\rho} \right) 
+ \left(\frac{\rho}{\ell}\right)^d (\partial_x^2 \delta g - \partial_t^2 \delta g) 
\biggr) \ .
\end{eqnarray}
As expected for a solution to the equations of motion at zeroth order, the first order term in the expansion is a total derivative.

\vskip 0.1in
\noindent
At second order, the expression is complicated but manageable:
\begin{eqnarray}
{\mathcal L}_2 &=& \frac{\sqrt{-g} }{2 \kappa^2}\biggl( -\frac{1}{2}(\partial_\mu \delta \phi) (\partial^\mu \delta \phi) + \frac{d^2-1}{\ell^2} e^{\gamma \phi} (\delta \phi)^2 + \frac{ e^{\gamma \phi} }{8\ell^2} \rho^2 (\partial_\rho \delta g^a_a)^2  \\
&& \hskip 0.5in
- \frac{d \gamma \rho}{2\ell^2} e^{\gamma \phi} \delta g^a_a \partial_\rho \delta \phi
+ \frac{\gamma d(d+1)}{2\ell^2} e^{\gamma \phi} \delta g^a_a \delta \phi
\biggr) \nonumber \ .
\end{eqnarray}
This restricted action reproduces the dynamical equations ${\rm EOM}_\varphi$ and ${\rm EOM}_{tt}$  in the scalar-trace sector, but not the constraint equations ${\rm EOM}_{t\rho}$ and ${\rm EOM}_{\rho \rho}$ associated with the fields that have been set to zero.
To remedy this potential issue, in evaluating the current below,  we use the solutions of ${\rm EOM}_{t\rho}$ and ${\rm EOM}_{\rho \rho}$.

\vskip 0.1in
\noindent
There is a subtle point here about whether one should trust a current derived from a quadratic action that does not reproduce the first-order constraints. The relevant observation is that the physical branch used below is obtained from the full linearized equations with $c_1 = c_2 = c_+ = 0$. 
On this branch $g_+ = g_2 = 0$, while $g_-$ and $\varphi$ obey the radial constraints ${\rm EOM}_{t\rho}$ and ${\rm EOM}_{\rho \rho}$.
Thus any term in the full symplectic current containing $g_+$, $g_2$, or their derivatives vanishes on the physical branch.
Even though we derive the current using a reduced action, it should correspond to the current derived from the full
second order action.

\vskip 0.1in
\noindent
Given this quadratic action, we may  compute a flux in the $\rho$, $t$, and $x$ directions.
We define the conjugate momenta
\begin{equation}
\Theta^\mu_{\phi} \equiv \frac{\partial {\mathcal L_2}}{\partial \partial_\mu \delta \phi} \ , \; \; \;
\Theta^\mu_{g} \equiv \frac{\partial {\mathcal L_2}}{\partial \partial_\mu \delta g} \ . 
\end{equation}  
We can then compute the flux
\begin{equation}
\Pi^\mu = i (\Theta^\mu_\phi \delta \phi + \Theta^\mu_g \delta g - c.c.)\ .
\end{equation}
Equivalently, this expression is the conserved Wronskian current evaluated on a complex mode and its conjugate.
We find
\begin{eqnarray}
\Pi^\rho &=&  \frac{1}{\kappa^2}\ell^d (\omega^2 - k^2)^{1/2} |c|^2 \ , \\
\Pi^t &=& \frac{1}{\kappa^2} \omega \ell^d  (\omega^2-k^2)^{1/2} \frac{\pi}{2}  \rho \left|H^{(1)}_{\frac{d-3}{2}}(\rho \sqrt{\omega^2-k^2})\right|^2 |c|^2 \ , \\
\Pi^x &=&\frac{1}{\kappa^2} k \ell^d (\omega^2-k^2)^{1/2} \frac{\pi}{2}   \rho \left|H^{(1)}_{\frac{d-3}{2}}(\rho \sqrt{\omega^2-k^2})\right|^2 |c|^2 \ .
\end{eqnarray}
For $d=2$ and $d=4$, the Hankel-function dependence cancels from the longitudinal flux components, 
leaving the same $(\omega , k)$-dependence as a flat-space plane wave:
\begin{eqnarray}
\Pi^t =  \frac{1}{\kappa^2} \ell^d  \omega |c|^2\ , \; \; \; \Pi^x =  \frac{1}{\kappa^2} \ell^d  k |c|^2\ .
\end{eqnarray}
For larger, even $d$, the flux in the $t$ and $x$ directions becomes polynomial in $\rho$ and is no longer the same as a plane wave in flat space.

\vskip 0.1in
\noindent
In the pure $AdS_3$  interface problem of ref.\ \cite{Bachas:2020yxv}, the scattered modes are boundary-graviton-like gauge modes. Their amplitudes directly determine the stress-tensor one-point functions on the two sides of the interface, and the bulk matching condition between the incident, reflected, and transmitted wave amplitudes $I = R+T$ has the immediate interpretation of energy-flux conservation in the dual CFT. The present Chamblin-Reall problem is different. For $d>1$, the scattered excitation is a physical propagating dilaton-graviton wave. The coefficients $I$, $R$, and $T$ are therefore coefficients of bulk wave solutions rather than directly normalized boundary energy fluxes. Moreover, the brane redistributes a single incident wave into a continuum of reflected, transmitted, and evanescent channels. For these reasons, we introduce a conserved bilinear flux for the physical bulk mode. This flux is the conserved wave flux in the scattering problem. Its precise decomposition into dual stress-tensor and scalar one-point functions requires a more complete holographic dictionary for these Chamblin-Reall backgrounds, which we leave for future work.

\section{Linearized scattering problem}
\label{sec:IJ}

The linearized matching problem has two stages. First, continuity of the induced metric and dilaton constrains the brane embedding functions and relates the incident, reflected, and transmitted amplitudes. Second, the Israel and dilaton normal-jump conditions impose additional compatibility relations. For $d=1$, the system can be solved with a single incoming, reflected and transmitted ray \cite{Bachas:2020yxv}. For $d>1$, a finite set of rays is insufficient: the interface redistributes the incident wave into a continuum of outgoing angles and evanescent surface modes, taking full advantage of the new physical bulk degrees of freedom that are not present in the $d=1$ case.

\vskip 0.1in
\noindent

A useful cartoon of the $d>1$ process is scattering from a rough translucent window: an incident beam is partly transmitted, partly reflected, and partly converted into evanescent surface modes.
The dilaton profile, which changes the effective tension of the brane as a function of the radial coordinate $\rho$, appears to be enough to produce the effect of roughness in the  interface, as we will see.

\vskip 0.1in
\noindent

To keep things simple at first, we allow for just a single incident angle of dilaton-graviton waves, a single reflected angle, and a single transmitted angle.  As the equations are linear, we can always take a linear superposition of the modes later on.  Eventually, we will want to sum over all angles, and even continue these angles into the complex plane in order to capture the evanescent modes.

\vskip 0.1in
\noindent

In the $d=2$ and $d=4$ cases which will be the focus here, the Hankel functions describing the dilaton-graviton waves 
reduce to simple exponentials, as we saw in  section \ref{sec:linearized}.   In the $x\rho$-coordinate system, we identify angles $\alpha_I$, $\alpha_R$, and $\alpha_T$ associated with the incoming, reflected and transmitted radiation respectively (see figure \ref{fig:basicsetup}).  Assuming a wave profile $e^{-i \omega t + i k x}$, the angles are defined through the wave vectors as
\begin{equation}
k_I = \omega \sin \alpha_I \ , \; \; \; k_R = - \omega \sin \alpha_R \ , \; \; \; k_T = \omega \sin \alpha_T  \ . 
\end{equation}

\vskip 0.1in
\noindent
We also include the gauge solution associated with the integration constant $c_+$ or equivalently the infinitesimal coordinate transformation 
$\omega v_{(1)} - k v_{(2)}$. 
While it is not a priori clear these gauge solutions need to have the same $k_I$, $k_R$, and $k_T$ as the physical waves, it turns out we do not need to introduce different wave vectors for these gauge solutions to impose our boundary conditions.  

\vskip 0.1in
\noindent
In deriving the linearized waves and gauge solutions in section \ref{sec:linearized}, 
we characterized them with the amplitudes $c$ and $c_+$.  To distinguish
the different types of waves, we now distinguish these constants through the following relabelling:
\begin{eqnarray}
c_+ &\to&(\ell_L)^{\frac{3d}{2}} I_G  \ , \; \; \; (\ell_L)^{\frac{3d}{2}}R_G \ , \; \; \; \mbox{or} \; \; \; (\ell_R)^{\frac{3d}{2}}T_G  \ , \\
c &\to& (\ell_L)^{-\frac{d}{2}} I  \ ,  \; \; \;
   (\ell_L)^{-\frac{d}{2}} R   \ , \; \; \; \mbox{or} \; \; \;
 (\ell_R)^{-\frac{d}{2}} T  \ ,
\end{eqnarray}
where the factors of $\ell_L$ and $\ell_R$ are added for later convenience.

\vskip 0.1in
\noindent
Let the coordinates on the interface be $(\tau, \sigma)$.  From the point of view of the left $(t_L, x_L, \rho_L)$ or right $(t_R, x_R, \rho_R)$ coordinate system, we can think of the location of the interface as a map from ${\mathbb R}^2 \to {\mathbb R}^3$.  The fluctuations will induce small corrections of order $\epsilon$ in the embedding map such that 
\begin{eqnarray}
t &=& \tau + \epsilon e^{-i \tau \omega} \lambda(\sigma) \ , \\
u &=&\epsilon e^{-i \tau \omega} \delta(\sigma) \ , \\
v &=& \sigma  + \epsilon e^{-i \tau \omega} \zeta(\sigma) \ .
\end{eqnarray}
We work in a rotated coordinate system where $\rho = u \cos \theta + v \sin \theta$ and $x = -v \cos \theta + u \sin \theta$ and the unperturbed interface sits at $u=0$.  The $\theta_{L/R}$ angles defined previously are $\theta = \frac{\pi}{2} + \theta_L$ and $\theta = \frac{\pi}{2} - \theta_R$ in terms of $\theta$.  
We can then insert $R/L$ subscripts on the quantities $t$, $x$, $\rho$, $\zeta$, $\lambda$ and $\delta$ as needed when referring to the right or 
left side of the interface.  Reparametrization invariance on the interface means that two of these six functions are superfluous.
Indeed, we will find the solution depends only on the differences $\zeta_L - \zeta_R$ and $\lambda_L - \lambda_R$.  

\vskip 0.1in
\noindent
To construct the extrinsic curvature, we take the cross product of the two vectors $\partial_\tau (t, u,v)$ and $\partial_\sigma (t, u,v)$.
This object naturally has a lower index.  From it, we can construct a unit normalized normal vector $n_\mu$.  The extrinsic curvature is then the covariant derivative $K_{\mu\nu} =  h^\lambda_\mu h^\rho_\nu \nabla_\lambda n_\rho$ projected onto the interface with $h_{\mu\nu} = g_{\mu\nu} - n_\mu n_\nu$.  Finding the continuity and junction conditions at linear order is then a tedious perturbative expansion. With a computer algebra package, the expansion can be done by brute force although details have been formalized in the literature (see for example \cite{Mukohyama:2000ga,Mars:2005ca}).  We review some of these details in appendix \ref{sec:linearization}.

\vskip 0.1in
\noindent
The solution strategy is as follows.  Continuity of the metric at the interface imposes 
\begin{align}
h_{\tau \tau}:& \; \; \; i\omega \lambda + d \frac{\Delta}{\sigma} =  \ldots \ , \\
h_{\tau \sigma}:& \; \; \;   \lambda' + i\omega \zeta = \ldots  \ , \\
h_{\sigma \sigma}:& \; \; \; \zeta' - d \frac{\Delta}{\sigma} = \ldots
\end{align}
where $\Delta = -\tan \theta_L \delta_L  - \tan \theta_R \delta_R + \zeta$, $\delta = \delta_L - \delta_R$, $\zeta = \zeta_L - \zeta_R$, $\lambda = \lambda_L - \lambda_R$.  
The ellipses on the right hand side are source terms that depend linearly on the reflection and transmission coefficients $I$, $R$, $T$, $I_G$, $R_G$, and $T_G$.  
Dependence on $\lambda_L + \lambda_R$ and $\zeta_R + \zeta_L$ drops out.  
From this system, we get a set of coupled first order ODEs for $\lambda$ and $\zeta$ and a way to solve algebraically for $\Delta$ from $\lambda$.  In the absence of sources, this system supports a homogeneous solution which corresponds to a plane wave on the brane.  

\vskip 0.1in
\noindent
In fact, there is a problem at this point.  If we also impose continuity of the dilaton, then we immediately get a solution for $\Delta$ which also turns off the homogeneous solution.  The brane, in the presence of the dilaton, is rigid.  More than that, we find two distinct ways to solve for $\zeta'$. Their compatibility gives a constraint on the amplitudes $I$, $R$ and $T$.  In the particular case $d=2$, that constraint is
\begin{eqnarray}
0 &=&
 + e^{-i \sigma \omega \cos(\alpha_I-\theta_L)} I 
\left( \sec \theta_L \cos \alpha_I- i \sigma \omega \sin^2 (\alpha_I-\theta_L)\right) \nonumber \\
&& + e^{i \sigma \omega \cos(\alpha_R+\theta_L)} R 
\left(\sec \theta_L  \cos \alpha_R  - i \sigma \omega \sin^2 (\alpha_R+\theta_L)   \right)  \nonumber \\
&&  -e^{i \sigma \omega \cos(\alpha_T+\theta_R)} T
\left(\sec \theta_R  \cos \alpha_T  - i \sigma \omega \sin^2 (\alpha_T+\theta_R) \right) \ . 
\label{constraint1}
\end{eqnarray}

\vskip 0.1in
\noindent
Next we move on to the Israel junction conditions.  The $K_{\tau \sigma}$ condition provides a solution for $\delta'(\sigma)$ in terms of sources.  
In the $d=1$ case, this relation essentially completes the matching problem.  The $K_{\tau \tau}$ condition, which can be reduced to an equation involving only $\delta''$ and $\delta'$ using metric continuity, is automatically satisfied.  The $K_{\sigma \sigma}$ condition, which can be reduced to an equation involving only $\delta'$ and $\delta$ using metric continuity, is satisfied provided $I = R+T$.  We return to the metric continuity relations and solve for $\zeta$ and $\lambda$.  Appropriate boundary conditions (waves moving purely into the geometry along the interface) then yield $T$ and $R$ independently.

\vskip 0.1in
\noindent
For $d>1$, there is another obstruction. 
We can again solve for $\delta'$ from $K_{\tau \sigma}$.  The $K_{\tau \tau}$ relation then provides a solution for $\lambda$.  
(The function $\lambda$ shows up in this Israel junction condition with coefficient $d-1$.)  Moving to $K_{\sigma \sigma}$, we solve for $\delta$, which now provides an  independent way of computing $\delta'$.  That consistency relation takes the following form in $d=2$:
\begin{eqnarray}
0&=&+ e^{-i \sigma \omega \cos (\alpha_I - \theta_L)}  \sin(\alpha_I - \theta_L) I 
-e^{i \sigma \omega \cos (\alpha_R + \theta_L)} \sin(\alpha_R + \theta_L) R
 \nonumber \\
&&
 - e^{-i \sigma \omega \cos (\alpha_T + \theta_R)} \sin(\alpha_T + \theta_R) T  \ .
\label{constraint2}
\end{eqnarray}

\vskip0.1in
\noindent
The key new obstruction for $d>1$ is that the matching equations produce two independent functional constraints. (In $d=2$, they are (\ref{constraint1}) and (\ref{constraint2}).)  A single reflected and a single transmitted ray do not provide enough freedom to satisfy both constraints for all $\sigma$.
To solve this system, we swap $T$ with $\int T(\alpha_T) \rd \alpha_T$, $R$ with $\int R(\alpha_R) \rd \alpha_R$, and $I$ with $\int I(\alpha_I) \rd \alpha_I$.  We have not been able to solve the general $d>1$ case, but we have solved the $d=2$ and $d=4$ cases, which are instructive.

\subsection{The $d=2$ case}

\begin{figure}
\begin{center}
a) \includegraphics[width=2.5in]{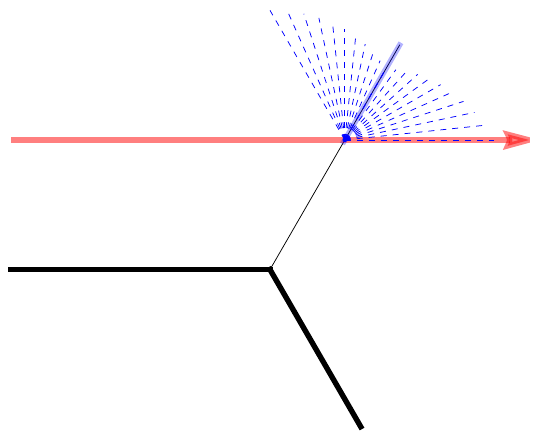}
b) \includegraphics[width=3in]{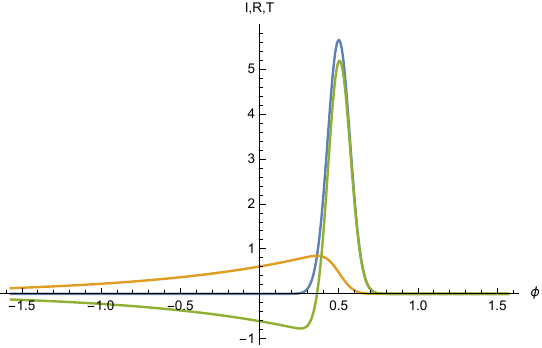}
\end{center}
\caption{
$d=2$ scattering: a) Radiation pattern in the $d=2$ case.  The solid blue segment along the interface indicates the evanescent surface contribution. 
b) Choosing $\tilde I(\myangle)$ to be a Gaussian wave packet (blue), the resultant transmitted $\tilde T(\myangle)$ (green) and reflected $\tilde R(\myangle)$ (orange)
distributions are plotted.
}
\label{fig:gaussian2d}
\end{figure}

The $d=2$ case provides a clean solution to this linearized scattering problem, with a clear choice of boundary conditions that leads to physics that is familiar both from a bulk and dual field theory perspective.
Promoting $T \to \int T(\alpha_T) \rd \alpha_T$, $R \to \int R(\alpha_R) \rd \alpha_R$, and $I \to \int I(\alpha_I) \rd \alpha_I$ converts the two constraints (\ref{constraint1}) and (\ref{constraint2}) to the following pair of integral equations:
 \begin{eqnarray}
0 &=&  + \int_{\theta_L}^{\theta_L+\pi} e^{-i \sigma \omega \cos(\alpha_I-\theta_L)} 
\left( \sec \theta_L \cos \alpha_I- i \sigma \omega \sin^2 (\alpha_I-\theta_L)\right)I(\alpha_I) \,  \rd \alpha_I \nonumber \\
&& + \int_{-\theta_L}^{\pi- \theta_L} e^{i \sigma \omega \cos(\alpha_R+\theta_L)} 
\left( \sec \theta_L \cos \alpha_R  - i \sigma \omega \sin^2 (\alpha_R+\theta_L)  \right) R(\alpha_R) \, \rd \alpha_R  \\
 && - \int_{-\theta_R}^{ -\theta_R+\pi} e^{i \sigma \omega \cos(\alpha_T+\theta_R)} 
\left( \sec \theta_R \cos \alpha_T  - i \sigma \omega \sin^2 (\alpha_T+\theta_R)   \right) T(\alpha_T) \,  \rd \alpha_T \ , \nonumber
\end{eqnarray}
\begin{eqnarray}
0&=&+ \int_{\theta_L}^{\theta_L+ \pi} e^{-i \sigma \omega \cos (\alpha_I - \theta_L)}  \sin(\alpha_I - \theta_L)   I (\alpha_I) \,  \rd \alpha_I \nonumber \\
&&
-\int_{-\theta_L}^{\pi-\theta_L} e^{i \sigma \omega \cos (\alpha_R + \theta_L)} \sin(\alpha_R + \theta_L)  R(\alpha_R)  \, \rd \alpha_R \nonumber \\
&&- \int_{-\theta_R}^{-\theta_R+\pi} e^{i \sigma \omega \cos (\alpha_T + \theta_R)} \sin(\alpha_T + \theta_R)    T(\alpha_T) \, \rd \alpha_T  \ .
\end{eqnarray}
We have not yet introduced an evanescent contribution to the solution.  Such an evanescent contribution could be included by continuing the integral into the complex plane.  We will return to this possibility below.

\vskip 0.1in
\noindent
We solve this system of integral equations by introducing a new common angle, $\myangle = \alpha_I-\theta_L - \frac{\pi}{2}$, $\myangle= - \frac{\pi}{2} + \alpha_R+\theta_L$, and $\myangle = -\frac{\pi}{2} + \alpha_T+\theta_R$, 
with respect to which all the plane wave factors become the same.  
This choice leads to the relation 
\begin{equation}
\alpha_I - \theta_L = \alpha_R + \theta_L \ ,
\end{equation}
 which is the condition that the angle of incidence equals the angle of reflection.  Similarly we have 
 \begin{equation} 
 \alpha_I -\theta_L = \alpha_T + \theta_R \ , 
 \end{equation}
  which means the angle of incidence equals the angle of transmission (relative to the geometry set up by the interface, not the boundary).  
Equivalently, the matching conditions enforce conservation of the wave-vector component parallel to the interface.

\vskip 0.1in
\noindent
Defining
$\tilde I(\myangle) \equiv I(\frac{\pi}{2} + \myangle+\theta_L) $, $\tilde R(\myangle) \equiv R(\frac{\pi}{2}+\myangle-\theta_L) $, and
$\tilde T(\myangle) \equiv T(\frac{\pi}{2} + \myangle-\theta_R) $, the pair of integral equations simplifies substantially to

\begin{eqnarray}
0 &=&  \int_{-\pi/2}^{\pi/2}  e^{-i \sigma \omega \sin \myangle} \biggr( \tilde I(\myangle) 
\left( \sin( \theta_L+ \myangle) \sec \theta_L + i \sigma \omega \cos^2 \myangle \right)  \nonumber \\
&& \hskip 1in - \tilde R(\myangle)  
\left( \sin (\theta_L-\myangle) \sec \theta_L - i \sigma \omega \cos^2 \myangle \right)   \nonumber \\
 && \hskip 1in +  \tilde T(\myangle) 
\left( \sin (\theta_R-\myangle) \sec \theta_R - i \sigma \omega \cos^2 \myangle \right)  \biggl) \rd \myangle \ , \\
0&=& \int_{-\pi/2}^{ \pi/2} e^{-i \sigma \omega \sin \myangle}   \cos \myangle  \biggl(   \tilde I(\myangle)  
-   \tilde R(\myangle)
-   \tilde T(\myangle)   \biggr) \rd \myangle \ .
\end{eqnarray}

The first constraint we can integrate by parts to put the $\sigma \omega$ dependence purely in the plane wave factor:
\begin{eqnarray}
\label{dtwointegral}
0 &=&  \int_{-\pi/2}^{\pi/2}  e^{-i \sigma \omega \sin \myangle}  \cos \myangle \biggr( 
\tilde I'(\myangle)   + \tilde R'(\myangle) - \tilde T'(\myangle) 
\nonumber \\
&& 
\hskip 0.5in + \tilde I(\myangle) \tan \theta_L - \tilde R(\myangle) \tan \theta_L+ \tilde T(\myangle) \tan \theta_R \biggr) \rd \myangle 
 \\
 && \hskip 0.5in - \left. (\tilde I(\myangle) + \tilde R(\myangle) - \tilde T(\myangle))\cos \myangle  \right|_{\myangle = -\frac{\pi}{2}}^{\myangle = \frac{\pi}{2}}
 \ , \nonumber
\end{eqnarray}
where we have included the boundary terms even though they nominally vanish at $\myangle = \pm \frac{\pi}{2}$.  

\vskip0.1in
\noindent
The integral constraints need to be true mode by Fourier mode, which converts these integral equations into an algebraic and a first order
differential constraint on $\tilde I(\myangle)$, $\tilde R(\myangle)$, and $\tilde T(\myangle)$.  
\begin{eqnarray}
\label{IRTreld2}
0 &=& \tilde I (\myangle) - \tilde R(\myangle) - \tilde T(\myangle) \ , \\
\label{IRTfirstorderd2}
0&=&  \tilde I'(\myangle) + \tilde R'(\myangle) - \tilde T'(\myangle) + \tilde I(\myangle) \tan \theta_L - \tilde R(\myangle) \tan \theta_L + \tilde T(\myangle) \tan \theta_R  \ . 
\end{eqnarray}
Remarkably, these relations have coefficients which are independent of $\myangle$.  This $\myangle$ independence persists for the $d=4$ case, but we were not able to find such a simple presentation of the relations for higher, even $d$.  The coefficients are also independent of $\omega$, 
which means the behavior of the reflection and transmission in this system is independent of the energy of the incoming rays.  
(For odd $d$, the Hankel functions do not simplify, and we have made little progress in solving the constraints.)

\vskip 0.1in
\noindent
Indeed, the constraints allow us to determine $\tilde R(\myangle)$ and 
$\tilde T(\myangle)$ in terms of $\tilde I(\myangle)$ up to a single integration constant $C$:
\begin{eqnarray}
\label{Rsol}
\tilde R(\myangle) &=& e^{\frac{m \myangle}{2}} \left( C +\frac{m}{2} \int_\myangle^{\frac{\pi}{2}} e^{-\frac{ms}{2}}
 \tilde I(s) \rd s \right) 
\ , \\
\label{Tanswer2d}
\tilde T(\myangle) &=& \tilde I(\myangle) - \tilde R(\myangle) 
\ . 
\end{eqnarray}
where we have defined $m \equiv \tan \theta_L + \tan \theta_R$.  Note that waves with $\myangle = \frac{\pi}{2}$ travel parallel to the interface and toward the $\rho=0$ boundary while waves with $\myangle = -\frac{\pi}{2}$ travel parallel but away from the boundary.  The AdS/CFT intuition here suggests choosing a ``causal'' boundary condition where the wave propagates into the geometry and away from the interface, thus setting $C=0$. 
(When we pass to the $d=4$ case, we will have to loosen this requirement.)  

\vskip 0.1in
\noindent
If we choose a distributional form for $\tilde I(\myangle) = \delta(\myangle - \alpha)$, such a choice leads to the following solution for $\tilde R(\myangle)$
with the $C=0$
boundary condition:
\begin{equation}
\label{Ranswer2d}
\tilde R(\myangle) =\frac{m}{2} e^{m\frac{\myangle-\alpha}{2}}\Theta(\alpha-\myangle) \ . 
\end{equation}
Note this boundary condition is choosing the exponentially damped solution, which seems physically reasonable in the cases where the tension may be very large and the exponential growth can get very big as $\myangle \to \frac{\pi}{2}$.  
It also zeroes out support for the reflected and transmitted amplitude for waves traveling into the boundary, which meshes with AdS/CFT intuition that causal boundary conditions are associated with waves propagating away from the boundary and into the interior of the geometry.  As mentioned above, the $d=4$ case will require
a more general boundary condition.

\vskip 0.1in
\noindent
A more careful take on boundary conditions requires looking at the evanescent waves and considering the total flux.
From the discussion in section \ref{sec:flux}, there should be a flux associated with these waves that is proportional to $|I|^2$, $|R|^2$ and $|T|^2$.  
Flux conservation would require
\begin{equation}
\int |\tilde I|^2 \rd \myangle - \int |\tilde R|^2 \rd \myangle- \int |\tilde T|^2 \rd \myangle = 0 \ .
\end{equation}
Using (\ref{IRTreld2}), the left hand side can be rewritten as
\begin{equation}
\label{fluxredd2}
\int \left[ \tilde R^* ( \tilde I - \tilde R ) + \tilde R (\tilde I^* - \tilde R^* ) \right] \rd \myangle  \ .
\end{equation}
On the other hand, combining (\ref{IRTreld2}) and (\ref{IRTfirstorderd2}) gives the single first order equation for just $\tilde R$ and $\tilde I$ (which is how we obtained (\ref{Rsol})):
\begin{equation}
\tilde R'(\myangle) =  (\tilde R(\myangle) - \tilde I(\myangle))m  \ .
\end{equation}
Thus the integral (\ref{fluxredd2}) is actually a total derivative:
\begin{equation}
-\frac{1}{m} \int \frac{\rd}{\rd\myangle} |\tilde R(\myangle)|^2 \rd \myangle \ .
\end{equation}
Provided $\tilde R(\myangle)$ vanishes at the endpoints, the flux is conserved.  While we chose our $\myangle = \frac{\pi}{2}$ boundary condition to set $\tilde R(\frac{\pi}{2}) =0$, in the other direction $\tilde R(-\frac{\pi}{2})$ does not vanish.  The amount by which flux fails to be conserved, in the delta function case, is
\begin{equation}
\frac{1}{m}  \left| \tilde R\left(-\frac{\pi}{2}\right) \right|^2 = \frac{m}{4} e^{-m (\alpha + \frac{\pi}{2})} \ , 
\end{equation}
which is exponentially small for positive tension interfaces but not zero.

\vskip 0.1in
\noindent
Where does this extra flux go?  
The endpoint term has a natural interpretation as flux carried by evanescent surface modes. These modes are obtained by continuing the angular variable into the complex plane $\myangle = \frac{\pi}{2}-i \chi$ or $\psi = -\frac{\pi}{2} + i \chi$, $\chi>0$, 
where the transverse momentum becomes imaginary.
Along these complex portions of the contour, the plane wave factor $-i \sigma \omega \sin \myangle = \mp i \sigma \omega \cosh \chi$ in the integral continues to be consistent with a wave traveling along the brane, with smaller and smaller wavelength as $\chi$ grows.  
 Further the branch with $\myangle = \frac{\pi}{2} - i \chi$ would be traveling toward the $\rho = 0$ 
 boundary and  $\myangle =- \frac{\pi}{2} + i \chi$ away from it.
 In the bulk, the reflected waves should have a component $e^{- i u \omega \cos \myangle}$ while the transmitted ones a factor of 
 $e^{ i u \omega \cos \myangle}$.
 Continuing onto either evanescent branch $\myangle = \mp \frac{\pi}{2} \pm i \chi$, the reflected waves will damp as $e^{u \omega \sinh \chi}$ as $u$ decreases from the interface location at zero, 
 while the transmitted ones damp as $e^{-u \omega \sinh \chi}$ as $u$ grows from zero. 
 Introducing a very small imaginary part to the exponential behavior $m \to m + i \epsilon$ will kill the reflected and transmitted amplitudes in the limit $\myangle \to -\frac{\pi}{2} + i \infty$ and forces us to choose the boundary conditions above so that
the solution will not blow up in the other limit $\myangle \to \frac{\pi}{2} - i \infty$.\footnote{%
There are two other potentially natural boundary conditions in this model.  We could have set $\tilde R(-\pi/2) = 0$.  We would find then that  $|\tilde R|^2 + |\tilde T|^2 > |\tilde I|^2$, which looks like stimulated emission by the interface, converting evanescent waves into traveling waves.  Also, we could tune the system such that $|\tilde R(\pi/2)| = |\tilde R(-\pi/2)|$ in which case the flux would balance.  Both of these cases involve surface modes traveling away from the singularity and toward the boundary.  This sensitivity to the physics of the singularity suggests they be discarded.
}

\vskip 0.1in
\noindent
The delta function source for $\tilde I$ is idealized.  To get a better sense of what the interface actually does, it is perhaps useful to look at a Gaussian source instead.  The result is plotted as figure \ref{fig:gaussian2d}.

\vskip 0.1in
\noindent
We focused above on the flux, which from a bulk perspective should be conserved.  Given the constraint (\ref{Tanswer2d}), $\tilde I = \tilde R + \tilde T$, we may try to give a physical interpretation of the amplitudes themselves, as was done in the $d=1$ case for pure $AdS_3$ \cite{Bachas:2020yxv} where a similar constraint was found and interpreted as energy conservation in the dual field theory. 
We can calculate the total reflected amplitude,
\begin{equation}
\int_{-\frac{\pi}{2}}^{\frac{\pi}{2}} \tilde R(\myangle) \rd \myangle = 1 - \exp \left( -\left(\frac{\pi}{2} +\alpha \right) \frac{m}{2} \right) \ ,
\end{equation}
and also the total transmitted amplitude,
\begin{equation}
\int_{-\frac{\pi}{2}}^{\frac{\pi}{2}} \tilde T(\myangle) \rd \myangle =  \exp \left( - \left(\frac{\pi}{2} + \alpha \right) \frac{m}{2} \right) \ .
\end{equation}
Keeping the tension positive $m >0$ 
means the exponent will always be negative in the range $-\frac{\pi}{2} < \alpha < \frac{\pi}{2}$
and hence the transmitted and reflected amplitudes will be between zero and one, which seems sensible.  Negative tension branes are unstable and may amplify the reflected and transmitted amplitudes, which here would correspond to magnitudes larger than one.  
Curiously the ratio of the transmitted and reflected amplitudes has the form of a Bose-Einstein-like distribution
\begin{equation}
\frac{\int_{-\frac{\pi}{2}}^{\frac{\pi}{2}} \tilde T(\myangle) \rd \myangle}{\int_{-\frac{\pi}{2}}^{\frac{\pi}{2}} \tilde R(\myangle) \rd \myangle}
= \frac{1}{e^{\left(\frac{\pi}{2}+\alpha \right)\frac{m}{2} } - 1 } \ .
\end{equation}
The process becomes purely transmissive in the limit $\alpha \to -\frac{\pi}{2}$, where the incoming beam of light is aimed almost tangent to the interface and away from the $\rho=0$ boundary.

\vskip 0.1in
\noindent
For completeness, we give the full solution for the embedding function $\lambda$, $\zeta$, and $\delta$ along with some details of the gauge modes in appendix \ref{sec:solns}.

\subsection{The $d=4$ case}
\label{sec:d4}
For this $d=4$ case, we will eventually be faced with two competing boundary-condition choices, one of which is appealing from an optics perspective but produces sensitivity to the singularity in the bulk spacetime, the other of which makes sense holographically but introduces an exponentially growing evanescent mode moving toward the singularity that does not satisfy the naive boundary conditions established in the course of the derivation below.  As we will argue, our preference is for the holographic choice which we believe can eventually be put on a firmer footing through introducing a black-brane horizon to the geometry, although we do not do that here.

\vskip 0.1in
\noindent
We take advantage of our experience with the $d=2$ case to employ  immediately the angle $\myangle$ and make the redefinitions
$\tilde I(\myangle) \equiv I(\frac{\pi}{2} + \myangle+\theta_L) $, $\tilde R(\myangle) \equiv R(\frac{\pi}{2}+\myangle-\theta_L) $, and
$\tilde T(\myangle) \equiv T(\frac{\pi}{2} + \myangle-\theta_R) $.
The integral constraints are
\begin{eqnarray}
\label{firsteqd4}
0&=& \int_{-\frac{\pi}{2}}^{\frac{\pi}{2}} \rd \myangle \, e^{-i \sigma \omega \sin \myangle} \biggl( ( \tan \theta_L + i \sigma \omega \cos \myangle ) \tilde I(\myangle) 
 + (\tan \theta_L - i \sigma \omega \cos \myangle) \tilde R(\myangle) \nonumber \\
&& \hskip 0.5in + (\tan \theta_R - i \sigma \omega \cos \myangle) \tilde T(\myangle) \biggr) \ , \\
\label{secondeqd4}
0 &=&  \int_{-\frac{\pi}{2}}^{\frac{\pi}{2}} \rd \myangle \, e^{-i \sigma \omega \sin \myangle} \biggl( \nonumber  \\
&&(-\sigma^2 \omega^2 \cos^2 \myangle + i \sigma \omega( \sin \myangle-3 \cos \myangle \tan \theta_L ) -2 + 3 \sec^2 \theta_L)\tilde I(\myangle) \nonumber \\
&&+(-\sigma^2 \omega^2 \cos^2 \myangle + i \sigma \omega(\sin \myangle-3 \cos \myangle \tan \theta_L ) - 2 +3\sec^2 \theta_L)\tilde R(\myangle) \nonumber \\
&&+(\sigma^2 \omega^2 \cos^2 \myangle - i \sigma \omega( \sin \myangle-3 \cos \myangle \tan\theta_R) + 2 -3\sec^2 \theta_R)\tilde T(\myangle) \biggr) \ .
\end{eqnarray}
By restricting to real $\myangle$, we are for the moment ignoring the contributions from the evanescent modes. 
As in the $d=2$ case we must integrate by parts.  The first integral equation leads to the first order differential equation
\begin{eqnarray}
\label{firstorder}
0 &=& \tilde I'(\myangle) -\tilde R'(\myangle)  - \tilde T'(\myangle) + \tan \theta_L \tilde I(\myangle) + \tan \theta_L \tilde R(\myangle) + \tan \theta_R \tilde T(\myangle) \nonumber \\
&& - (\tilde I(\myangle) - \tilde R(\myangle) -\tilde T(\myangle) )
 \left( \delta\left(\myangle - \myangle_f \right) - \delta\left(\myangle -\myangle_i \right) \right) \ ,
\end{eqnarray}
where we have converted the boundary terms created by integration by parts into Dirac delta function contributions to the differential equation,
where $\myangle_i = -\frac{\pi}{2}$ and $\myangle_f = \frac{\pi}{2}$.  
The second integral constraint requires several integrations by parts:
\begin{eqnarray}
\label{secondorder}
0&=& \tilde I''(\myangle) + \tilde R''(\myangle) - \tilde T''(\myangle) + 3 \tan \theta_L \tilde I'(\myangle) -3 \tan \theta_L \tilde R'(\myangle) + 3 \cot \theta_R \tilde T'(\myangle) 
\nonumber \\
&&
+(-2 +3\sec^2 \theta_L) \tilde I(\myangle) + (-2+3 \sec^2 \theta_L) \tilde R(\myangle)
+(2 - 3\sec^2 \theta_R) \tilde T(\myangle)  \nonumber \\
&& - \biggl( \left(\tilde I'(\myangle) + \tilde R'(\myangle) - \tilde T'(\myangle) + 3 \tan \theta_L \tilde I(\myangle) - 3 \tan \theta_L \tilde R(\myangle) + 3 \tan \theta_R \tilde T(\myangle)\right)\nonumber \\
&&- i \sigma \omega \cos \myangle (\tilde I(\myangle) + \tilde R(\myangle) - \tilde T(\myangle))\biggr)  \left( \delta\left(\myangle - \myangle_f \right) - \delta\left(\myangle -\myangle_i \right) \right)  \ .
\end{eqnarray}
Here the evanescent waves will play an important role in making sure these boundary terms vanish, once we extend the $\myangle$ contour into the complex plain.  
In extending the contour, we will push $\myangle_i$ and $\myangle_f$ past $-\frac{\pi}{2}$ and $\frac{\pi}{2}$ respectively and into the complex plane.

\vskip 0.1in
\noindent
To solve the differential equations, we reduce them 
 to a single third order equation for either one of the amplitudes:
\begin{eqnarray}
\label{ETdiffeq}
E_T &\equiv& D_{(3)} \tilde T(\myangle)-
\tilde I'''(\myangle) - \tilde I'(\myangle)  = 0 \ , \\
\label{ERdiffeq}
E_R &\equiv& D_{(3)} \tilde R(\myangle) 
+m \tilde I''(\myangle) - \frac{3}{2} m p \tilde I'(\myangle) - \frac{m}{8}(4-3p^2+3m^2) \tilde I(\myangle)  = 0\ ,
\end{eqnarray}
where
\begin{equation}
D_{(3)}  \equiv (\partial_\myangle)^3 - 2m (\partial_\myangle)^2 + \left(\frac{3 m^2}{2} + 1 \right)  \partial_\myangle - \frac{1}{8} m (4 - 3p^2 + 3 m^2) \ .
\end{equation}
We have defined $m = \tan \theta_R + \tan \theta_L$ as before and $p = \tan \theta_R - \tan\theta_L$.
There are homogeneous solutions to this third order equation $ e^{\kappa \myangle}$ where $\kappa$ is a root of the cubic polynomial
\begin{equation}
\label{mycubic}
Q(\kappa) = \kappa^3 - 2 m \kappa^2 +  \left(\frac{3 m^2}{2} + 1 \right) \kappa - \frac{1}{8} m (4 - 3p^2 + 3 m^2)\ .
\end{equation}
Using these homogeneous solutions, we can construct a Green's function to get a general solution.
The behavior of these roots will be important in establishing the physical properties of the solution, and we write them explicitly in 
appendix \ref{sec:cubic}.

\vskip 0.1in
\noindent
The general idea is as follows.  Consider a third order inhomogeneous ODE with constant coefficients and a source term $s(\myangle)$:
\begin{equation}
y'''(\myangle) + a y''(\myangle) + b y'(\myangle) + c y(\myangle) = s(\myangle) \ .
\end{equation}
To this ODE, we associate the cubic polynomial $Q(\myangle) = \myangle^3 + a \myangle^2 + b \myangle + c = \prod_{i=1}^3 (\myangle- \kappa_i)$.  
We have the following identities:
\begin{equation}
\label{Qids}
\sum_{i=1}^3 \frac{1}{Q'(\kappa_i)} = 0 \ , \; \; \; \sum_{i=1}^3 \frac{\kappa_i}{Q'(\kappa_i)} = 0 \ , \; \; \; \sum_{i=1}^3 \frac{\kappa_i^2}{Q'(\kappa_i)} = 1\,,~~~
\sum_{i=1}^3 \frac{\kappa_i^3}{Q'(\kappa_i)} = \sum_{i=1}^3 \kappa_i
\ .
\end{equation}
From these identities, the general solution 
\begin{equation}
y(\myangle) = \sum_{i=1}^3 e^{\kappa_i \myangle} \left(a_i + \int^\myangle \frac{s(\varphi)}{Q'(\kappa_i)} e^{-\kappa_i \varphi} \rd \varphi \right) \ .
\end{equation}
follows, with integration constants $a_i$.

\vskip 0.1in
\noindent
For the transmitted and reflected amplitudes, we have then
\begin{eqnarray}
\tilde T(\myangle) &=& \sum_{i=1}^3 e^{\kappa_i \myangle}\left( t_i +  \int^\myangle \frac{\tilde I'''(\varphi) + \tilde I'(\varphi)}{Q'(\kappa_i)} e^{-\kappa_i \varphi} \rd \varphi \right) \ , \\
\tilde R(\myangle) &=& \sum_{i=1}^3 e^{\kappa_i \myangle} \left( r_i + \int^\myangle \frac{m((4 - 3 p^2 + 3 m^2) \tilde I(\varphi) + 12 p \tilde I'(\myangle) - 8 \tilde I''(\varphi))}{8 Q'(\kappa_i)} e^{-\kappa_i \varphi} \rd \varphi \right) \ .
\end{eqnarray}
Integrating by parts, we can rewrite these expressions as
\begin{eqnarray}
\tilde T(\myangle) &=& \sum_{i=1}^3 e^{\kappa_i \myangle}\left( t_i +  \int^\myangle  \tilde I(\varphi) \frac{A(\kappa_i)}{Q'(\kappa_i)} e^{-\kappa_i \varphi} \rd \varphi \right) 
+ \tilde I(\myangle) \ , \\
\tilde R(\myangle) &=& \sum_{i=1}^3 e^{\kappa_i \myangle} \left( r_i + \int^\myangle \tilde I(\varphi)  \frac{B(\kappa_i)}{Q'(\kappa_i)} e^{-\kappa_i \varphi}  \rd \varphi \right) \ .
\end{eqnarray}
where we have defined the auxiliary polynomials:
\begin{eqnarray}
A(\kappa) &\equiv& \kappa^3 + \kappa \ , \; \; \;
B(\kappa) \equiv \frac{m}{8} ((4 - 3 p^2 + 3 m^2) + 12 p \kappa - 8 \kappa^2) \ .
\end{eqnarray}

The boundary terms for the upper limits can mostly be ignored because of the identities (\ref{Qids}).  The boundary terms from the lower limits can be absorbed into the integration constants $t_i$ and $r_i$.
These auxiliary polynomials obey the identity
\begin{equation}
A(\kappa) \left( \kappa - \frac{1}{2}(p+m) \right) + B(\kappa) \left( \kappa + \frac{1}{2}(p-m) \right) = \left(\kappa - \frac{1}{2}(p-m) \right) Q(\kappa) \ .
\end{equation}
In particular, evaluated at a root,
\begin{equation}
\label{weirdid}
A(\kappa_i) \left( \kappa_i - \frac{1}{2}(p+m) \right) + B(\kappa_i) \left( \kappa_i + \frac{1}{2}(p-m) \right) = 0 \ ,
\end{equation}
the combination vanishes.

\vskip 0.1in
\noindent
Of course we don't really have six integration constants here.  The first order differential equation (\ref{firstorder}) constrains half the $r_i$ and $t_i$ in terms of the others.  Using the identity (\ref{weirdid}), 
the first order equation (\ref{firstorder}) reduces from
\begin{equation}
\tilde I' + \frac{1}{2}(m-p) \tilde I = \tilde T' - \frac{1}{2}(p+m) \tilde T +\tilde R' +  \frac{1}{2}(p-m) \tilde R
\end{equation}
to
\begin{equation}
0 =  \sum_{i=1}^3 e^{\kappa_i \myangle} \left[ \left( \kappa_i -\frac{1}{2}(p+m) \right) t_i +\left( \kappa_i +\frac{1}{2}(p-m) \right) r_i \right] \ .
\end{equation}
Thus we find that the weighted sum of $t_i$ and $r_i$ must vanish in the same way that the combination of $A(\kappa_i)$ and $B(\kappa_i)$ in (\ref{weirdid}) vanishes, 
\[
\left( \kappa_i -\frac{1}{2}(p+m) \right) t_i  = - \left( \kappa_i +\frac{1}{2}(p-m) \right) r_i  \ .
\]  

\vskip 0.1in
\noindent
If we take $\tilde I(\myangle) = \delta(\myangle - \alpha)$ to be a Dirac delta function, then
\begin{eqnarray}
\tilde T(\myangle) &=& \delta(\alpha-\myangle) + \sum_{i=1}^3 \left( \frac{A(\kappa_i)}{Q'(\kappa_i)} e^{\kappa_i (\myangle-\alpha)} \Theta(\myangle - \alpha) + t_i e^{\kappa_i \myangle} \right)  \ , \\
\tilde R(\myangle) &=&  \sum_{i=1}^3 \left( \frac{B(\kappa_i)}{Q'(\kappa_i)} e^{\kappa_i (\myangle-\alpha)} \Theta(\myangle - \alpha) + r_i e^{\kappa_i \myangle} \right) \ .
\end{eqnarray}
In the $d=4$ case, the evanescent rays are important.  If we were to treat the boundary conditions corresponding to the Dirac delta function terms in (\ref{firstorder}) and (\ref{secondorder}) literally, we would have too many boundary conditions (four) for the number of integration constants (three) at $\myangle = \pm \frac{\pi}{2}$.  By including the evanescent modes, we can push the boundary conditions off to $\myangle = \pm \frac{\pi}{2} \mp i \infty$ where the boundary conditions can be satisfied by arranging for the more generic condition that the transmitted and reflected amplitudes die to zero.  Let us call this choice the ``optics boundary condition''.

\vskip 0.1in
\noindent
In more detail, 
for any real $\kappa_i$, we follow the same strategy as in the $d=2$ case, adding a small imaginary part $i \epsilon$ such that the mode falls to zero exponentially in the limit $\myangle \to - \frac{\pi}{2} + i \infty$ and we choose the integration constant such that the mode is exactly zero for $\myangle > \alpha$ and along the $\myangle = \frac{\pi}{2} - i \chi$ branch.  
We know the complex $\kappa_i$ roots will come in complex conjugate pairs because the cubic equation is real.  We arrange the integration constants such that the one with positive imaginary part persists along the $-\frac{\pi}{2} + i \chi$ contour and dies to zero exponentially as $\chi \to \infty$.
Correspondingly, the one with negative imaginary part naively should be zeroed out on the $\myangle < \alpha$ part of the contour 
but allowed to persist along the other half $\frac{\pi}{2} > \myangle > \alpha$ and 
$\myangle = \frac{\pi}{2} - i \chi$ where it too will eventually die off exponentially to zero.  
We will see in a moment that this naive prescription for $\kappa_i$ introduces some problems.
A curious aspect of this setup is that depending on the precise properties of the roots $\kappa_i$, the reflected and transmitted amplitudes may grow exponentially for a brief period along the real part of the contour $-\frac{\pi}{2} < \myangle < \frac{\pi}{2}$, leading effectively to something that resembles stimulated emission by the interface.  The incoming radiation produces an excess of transmitted and reflected radiation.

\vskip 0.1in
\noindent
If we need to set the $\kappa_i$ mode to zero for $\myangle > \alpha$, we choose $r_i = 0 = t_i$.
If on the other hand we require the mode to vanish for $\myangle < \alpha$, then we can set
\begin{equation}
t_i = -\frac{A(\kappa_i)}{Q'(\kappa_i)} e^{-\kappa_i \alpha}
\end{equation}
to ensure $\tilde T(\myangle)$ contribution vanishes here.
The identity (\ref{weirdid}) then implies
\begin{equation}
r_i =  -\frac{B(\kappa_i)}{Q'(\kappa_i)} e^{-\kappa_i \alpha}
\end{equation}
and correspondingly that the $\tilde R(\myangle)$ contribution also vanishes in this region.  
In the case when the mode vanishes for $\myangle < \alpha$, the mode contribution to $\tilde R$ and $\tilde T$ can be written more compactly as  
\begin{equation}
- \frac{A(\kappa_i)}{Q'(\kappa_i)} e^{\kappa_i (\myangle-\alpha)} \Theta(\alpha-\myangle)  \; \; \; \mbox{or} \; \; \;
 - \frac{B(\kappa_i)}{Q'(\kappa_i)} e^{\kappa_i (\myangle-\alpha)} \Theta(\alpha-\myangle)  \ .
\end{equation}

\vskip 0.1in
\noindent
Let us try and understand the qualitative behavior of the roots as a function of $m$ and $p$.
Note 
\[
Q'(\kappa) = 3 \kappa^2 - 4 m \kappa + \left(\frac{3 m^2}{2} + 1 \right) \ .
\]
The determinant of this quadratic is $-2 m^2 -12 < 0$.  Thus the quadratic has no real roots, and we can conclude from the behavior at large $\kappa$ that $Q'(\kappa) > 0$.  Thus $Q(\kappa)$ can have at most one real root.  
If we shift the variable $\kappa = x + m/2$, we find the new polynomial
\[
x \left( x^2 - \frac{m}{2} x + \frac{1}{4}(4+m^2) \right) + \frac{3 p^2 m}{8} \ .
\]
The quadratic $x^2 - \frac{m}{2} x + \frac{1}{4}(4+m^2) $ also has a negative discriminant $-4 - \frac{3 m^2}{4}$
and hence no real roots.  Thus it's always positive, as is $3 p^2 m/ 8$.  The allowed roots of this shifted
cubic must be $x_* \leq 0$.  The maximum value $x_*=0$ is attained when $p=0$.  (We insist that $m>0$.)
In other words, the real root $\kappa_* < \frac{m}{2}$.  For the complex roots, we have that $\kappa_* + 2 {\rm Re}(\kappa_\pm) = 2m$.  Thus ${\rm Re}(\kappa_\pm) > \frac{3m}{4}$. 
As we show in appendix \ref{sec:EA},  $4-3p^2 + 3m^2 = 4 + 12 \tan \theta_L \tan \theta_R$ must be positive for stability of the effective potential.  Thus we conclude that $\kappa_* > 0$ as well since the product of the three roots must be positive.

\begin{figure}
\begin{center}
a) \includegraphics[width=2.5in]{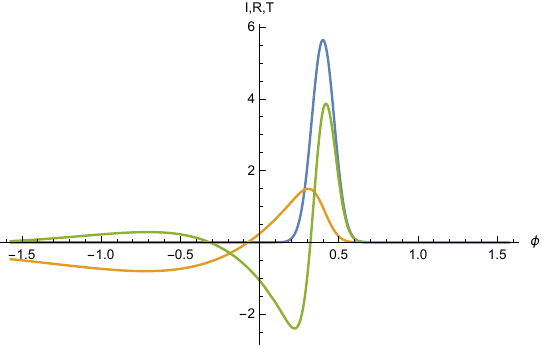}
b) \includegraphics[width=2.5in]{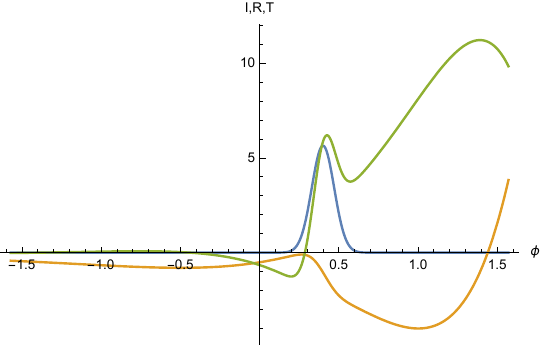}
\end{center}
\caption{
Choosing $\tilde I(\myangle)$ to be a Gaussian wave packet (blue), the resultant transmitted $\tilde T(\myangle)$ (green) and reflected $\tilde R(\myangle)$ (orange)
distributions are plotted in the $d=4$ case.
(a) holographic boundary conditions; (b) optics boundary conditions.
}
\label{fig:gaussian4d}
\end{figure}

\vskip 0.1in
\noindent
Putting this new information together with the ``optics'' boundary conditions above, we see that ${\rm Re}(\kappa_\pm)>0$  and ${\rm Re}(\kappa_*)>0$ are always positive.  
The proposed solution has the form
\begin{eqnarray}
\tilde T(\myangle) &=& \delta(\alpha-\myangle) \\
&& -\left( \frac{A(\kappa_*)}{Q'(\kappa_*)} e^{\kappa_* (\myangle-\alpha)} +\frac{A(\kappa_+)}{Q'(\kappa_+)} e^{\kappa_+ (\myangle-\alpha)} \right) \Theta( \alpha-\myangle)  + \frac{A(\kappa_-)}{Q'(\kappa_-)} e^{\kappa_- (\myangle-\alpha)} \Theta( \myangle - \alpha)  \ , \nonumber \\
\tilde R(\myangle) &=& - \left(\frac{B(\kappa_*)}{Q'(\kappa_*)} e^{\kappa_* (\myangle-\alpha)}+ \frac{B(\kappa_+)}{Q'(\kappa_+)} e^{\kappa_+ (\myangle-\alpha)} \right) \Theta( \alpha-\myangle)  + \frac{B(\kappa_-)}{Q'(\kappa_-)} e^{\kappa_- (\myangle-\alpha)} \Theta(\myangle -  \alpha)   \ .
\end{eqnarray}
Here we give the real root a small, positive imaginary part $\kappa_* \to \kappa_* + i \epsilon$ such that its amplitude dies out on the evanescent branch $\myangle = -\frac{\pi}{2} + i \chi$ of the solution.  The complex root $\kappa_+$ naturally dies there as well.  However,
the complex conjugate root $\kappa_-$ has negative imaginary part and naively should be extended along the $\myangle = \frac{\pi}{2} - i \chi$ 
branch, corresponding to an evanescent surface mode traveling in from the singularity and toward the boundary at $\rho = 0$.  
While consistent with the boundary conditions that emerged in our derivation and that 
we wrote as contact terms   in (\ref{firstorder}) and (\ref{secondorder}), the result is that the incident light hitting the interface appears to be amplified by flux coming in from the singularity.  Given an incoming Gaussian distribution, the reflected and transmitted amplitudes are plotted in figure \ref{fig:gaussian4d}b and are clearly larger in amplitude than the incoming radiation.  
These ``optics'' boundary conditions are sensitive to the presence of the singularity in the geometry and should be treated with caution and perhaps discarded.

\vskip 0.1in
\noindent
There is an alternative way to set boundary conditions which produces a more physical looking plot (figure \ref{fig:gaussian4d}a) where the 
thin brane absorbs flux from the incident radiation and the evanescent rays are all ingoing, away from the $\rho=0$ boundary.
We call these boundary conditions ``holographic''.
For a delta function $\tilde I(\myangle)$, the solution then takes the more compact form
\begin{eqnarray}
\label{Tanswer4d}
\tilde T(\myangle) &=& \delta(\alpha-\myangle) - \sum_{i} \frac{A(\kappa_i)}{Q'(\kappa_i)} e^{\kappa_i (\myangle-\alpha)} \Theta( \alpha-\myangle)  
 \ , \\
 \label{Ranswer4d}
\tilde R(\myangle) &=& - \sum_{i} \frac{B(\kappa_i)}{Q'(\kappa_i)} e^{\kappa_i (\myangle-\alpha)}  \Theta( \alpha-\myangle)   \ .
\end{eqnarray}
This type of ingoing boundary condition is familiar from the AdS/CFT literature where it produces causal Green's functions.
This choice, however, also has a problem.  If we let the $\kappa_-$ root persist along the $\myangle = -\frac{\pi}{2} + i \chi$ branch instead of the $\myangle = \frac{\pi}{2}-i \chi$ branch, its amplitude does not die off but instead grows exponentially with $\chi$.  
In other words, we are not satisfying for this mode the natural boundary conditions  (\ref{firstorder}) and (\ref{secondorder}) that emerged from our derivation.  We suspect the correct interpretation here is to imagine that we have regulated the singularity by introducing a very small black-brane horizon.  This horizon then enforces a purely ingoing boundary condition that removes the boundary constraints in (\ref{firstorder}) and (\ref{secondorder}).
This exponential growth is reminiscent of the radial growth of quasinormal-mode wavefunctions near a horizon, where the physically relevant condition is ingoing behavior rather than pointwise decay of the mode function.

\vskip 0.1in
\noindent
In summary, the $d=2$ problem gives a controlled scattering solution in which the interface redistributes a localized incident wave into a diffuse reflected/transmitted distribution together with evanescent surface modes. In $d=4$, the same matching equations expose a sensitivity to the singularity: enforcing strict evanescent damping leads to amplification sourced from the singularity, while imposing ingoing behavior toward a regulated horizon gives a more physical redistribution but requires relaxing the zero-temperature endpoint conditions.

\vskip 0.1in
\noindent
The full solution for the embedding functions can be found in appendix \ref{sec:solns}.
Before proceeding to the Discussion, we first demonstrate flux conservation in the $d=4$ case, which is substantially more involved than for $d=2$.  For completeness, we also discuss integrated amplitudes, although in absence of a $\tilde I = \tilde R + \tilde T$ condition, the relevance of these integrated quantities is not clear.

\subsubsection*{Flux Conservation for $d=4$}
The flux conservation relation on $\tilde I^2 - \tilde R^2 - \tilde T^2$ 
is substantially more intricate in this case, at least in the form we have derived it.  
As we did in the $d=2$ case, the first step is to replace $\tilde T(\myangle)$ with $\tilde R(\myangle)$ and $\tilde I(\myangle)$.  
The two differential equations (\ref{firstorder}) and (\ref{secondorder}) can be reassembled to show that
\begin{equation}
\tilde T(\myangle) = \frac{ (4 + 5 m^2-6mp+p^2)(\tilde I + \tilde R) + 8 m \tilde I' +(4p-12m) \tilde R' + 8 \tilde R''}{4+m^2+2mp+p^2} \ .
\end{equation}
Inserting this expression into $\tilde I^2 - \tilde R^2 - \tilde T^2$, we then try to express the result as a total derivative up to the 
equation of motion for $\tilde R$,
\begin{equation}
\frac{\rd F}{\rd \myangle} + \Lambda E_R \ ,
\end{equation}
where $E_R$ is the third order differential equation (\ref{ERdiffeq}) satisfied by $\tilde R$.  
Let us define the vector of amplitudes $V_i = (\tilde R, \tilde I, \tilde R', \tilde I', \tilde R'')$.  We claim there are $F$ and $\Lambda$  of the form
\begin{eqnarray}
F &=& \sum_{i \leq j} a_{ij} V_i V_j \ , \; \; \; \Lambda = \sum_i b_i V_i
\end{eqnarray}
where $a_{ij}$ and $b_j$ are $\myangle$ independent constants such that 
\begin{equation}
\tilde I^2 - \tilde R^2 - \tilde T^2 = \frac{\rd F}{\rd \myangle} + \Lambda E_R \ .
\end{equation}
The resulting expressions for the $a_{ij}$ and $b_j$ are messy.

\subsubsection*{Integrated amplitudes}

The total transmitted and reflected amplitudes are obtained by integrating over $\myangle$.  
Using the holographic boundary conditions, the non-evanescent pieces of the amplitude integrate to 
\begin{eqnarray}
\int_{-\frac{\pi}{2}}^{\frac{\pi}{2}} \tilde T(\myangle) d \myangle &=&  \sum_{i=1}^3  \frac{A(\kappa_i)}{\kappa_i Q'(\kappa_i)} e^{\kappa_i (-\frac{\pi}{2}-\alpha)}   \ , \\
\int_{-\frac{\pi}{2}}^{\frac{\pi}{2}} \tilde R(\myangle) d \myangle &=&  -1+ \sum_{i=1}^3 \frac{B(\kappa_i)}{ \kappa_i Q'(\kappa_i)} e^{\kappa_i(-\frac{\pi}{2} - \alpha)}  \ .
\end{eqnarray}
Note
\[
 \sum_{i=1}^3  \frac{A(\kappa_i)}{\kappa_i Q'(\kappa_i)} =\sum_{i=1}^3  \frac{B(\kappa_i)}{\kappa_i Q'(\kappa_i)} = 1 \ .
\]

\section{Discussion}

We have analyzed the linearized scattering of dilaton-graviton waves from a thin Chamblin-Reall brane. For $d=2$, the problem is controlled:
scattering resdistributes an incident wave into reflected, transmitted, and evanescent components. For $d=4$, the same matching problem can be solved formally, but the answer is sensitive to the boundary condition at the spacetime singularity.
The gravitational process shares some features with light scattering off a rough, translucent window. 
 A direct, transmitted component remains, while the rest of the incident ray is converted into diffuse reflected and transmitted radiation as well as evanescent waves traveling along the interface.
In particular, see
(\ref{Tanswer2d}), (\ref{Ranswer2d}), (\ref{Tanswer4d}), and (\ref{Ranswer4d}) for the precise form of the reflected and transmitted radiation.

\vskip 0.1in
\noindent
The cleanest interpretation is obtained in the $d=2$ case: the reflected and transmitted radiation is distributed over a cone of angles between the specularly reflected ray and the directly transmitted ray (see fig.~\ref{fig:gaussian2d}a). The $d=4$ case appears qualitatively similar, but is sensitive to boundary conditions at the singularity. The optics boundary condition, which demands exponential damping of evanescent amplitudes at large wave vector, forces some surface modes to propagate from the singularity toward the boundary. The resulting amplification of outgoing flux suggests that this prescription is sensitive to, and probably contaminated by, the singularity. The holographic boundary condition, in which all modes propagate away from the $\rho=0$ boundary and toward the singularity, avoids this flux amplification and gives a radiation pattern closer to the $d=2$ result (see fig.~\ref{fig:gaussian4d}a). Its cost is an exponentially growing evanescent amplitude. We interpret this growth as an artifact of imposing an ingoing condition in a zero-temperature singular geometry, and expect that a small black-brane horizon would provide the appropriate regulator.

\vskip 0.1in
\noindent
While the holographic dictionary is less certain for these Chamblin-Reall spacetimes than for AdS, 
we can deduce qualitative features
of the scattering in the dual field theories.  Given that $\theta_L$, $\theta_R >0$, and focusing on the $d=2$ case where the choice of boundary conditions was unambiguous,
the scattering produces something resembling thermalization.  
Conventionally, the radial coordinate $\rho$ should be a measure of renormalization group scale.
If $\alpha_I = \frac{\pi}{2}$, then the incoming radiation travels parallel to the $\rho=0$ boundary and 
can be associated with a fixed RG scale.  However, after the scattering process, all of the radiation -- the reflected, transmitted and evanescent radiation -- heads away from the boundary and toward $\rho \to \infty$, which should correspond to the infrared and a decrease in RG scale.  The heuristic picture is of radiation relaxing toward the infrared after the scattering event, although a precise statement would require a sharper holographic dictionary.

\vskip 0.1in
\noindent
The physics is very different from the behavior in the $d=1$ (pure $AdS_3$) case \cite{Bachas:2020yxv}.  
There, the field theory dual and the interface preserve conformal symmetry.  The modes excited in the scattering process are boundary gravitons and have no ability to propagate into the bulk.  They hug the boundary, and there is no analogous bulk channel into which they can dissipate.  
Like in our $d=2$ case, the amplitudes satisfy a conservation-like constraint condition $\tilde I = \tilde T + \tilde R$, but there is no need to take a sum over angles -- the modes all travel parallel to the boundary.
Instead of evanescent modes, the thin brane in ref.\  \cite{Bachas:2020yxv} is less rigid in the conformal case and supports a massless bending mode.

\vskip 0.1in
\noindent
This work opens up a number of avenues for future research.  The most immediate challenge is to resolve the boundary condition issues in the $d=4$ case, where the most promising direction is the addition of a black-brane horizon.  
More broadly, the $d=4$ analysis raises a classical boundary-condition problem for evanescent gravitational modes in singular spacetimes.
Although the boundary conditions were never at issue for the pure $AdS_3$ case, a black-brane was already considered there for other reasons \cite{Bachas:2021tnp,Bachas:2021fqo}.   Black-branes open up additional opportunities, allowing us to consider the effects of temperature and energy current flows in the dual field theory.  Instead of shooting gravitons at the interface, one may consider steady state currents and their associated shocks and rarefaction waves, like in  \cite{Bhaseen:2013ypa,Doyon:2014qsa,Lucas:2015hnv,Spillane:2015daa,Herzog:2016hob}.

\vskip 0.1in
\noindent
Although the analytic simplifications used here occur for even $d$, physically interesting uplifts include both $d=2$ and $d=3$, corresponding to three- and four-dimensional CFTs respectively.  (Note however that the $d=3$ case likely requires solving integral equations directly rather than reducing them to finite-order ODEs.) 
Finding explicit uplifted $AdS_4$ and $AdS_5$ geometries could cure the IR singularities at $\rho \to \infty$ and also provide a more straightforward field theory interpretation, through the canonical AdS/CFT dictionary, as scattering in three and four spacetime dimensional CFTs.  Additionally a higher-dimensional perspective would allow us to consider more general scattering processes, where the waves impinge on the interface at an angle in the field theory directions.  (The angles in our work are in the $\rho x$-plane, which has an interpretation as RG flow, but is less straightforward to interpret in the field theory.)  Alternately, one may try to glue our hyperscaling violating geometries to an asymptotically $AdS_3$ region like in \cite{Betzios:2018kwn}, in order to be able to invoke the standard AdS/CFT dictionary and give the results here a firmer field theory interpretation.

\vskip 0.1in
\noindent
There are also straightforward technical generalizations of our work.  Indeed,
we did not fully characterize this linearized scattering process for all values of the parameter 
$d$.   In the $d=2$ and $d=4$ cases,
the linearized fluctuations took a particularly nice, plane wave form.  
The most obvious and perhaps easiest extension of this work is to $d=6$ and higher even
dimensional cases, where the Hankel functions reduce to plane waves times polynomials
in the radial coordinate instead of just plane waves.  We looked 
at the $d=6$ and $d=8$ cases in the hopes of being able to provide a general even
$d$ solution.  As expected, the scattering process is governed by two differential equations for the scattering amplitudes
$\tilde I$, $\tilde R$ and $\tilde T$,
like (\ref{IRTreld2}) and (\ref{IRTfirstorderd2}) in the $d=2$ case or (\ref{firstorder}) and (\ref{secondorder}) in the $d=4$ case.
The differential equations appear to follow a pattern where their order is $\left(\frac{d}{2}-1, \frac{d}{2} \right)$. 
A crucial difference however is that the coefficients are no longer constant and instead depend on the angle, at least in the forms
that we explored, making them more challenging to solve. 
This structural change for the constraining differential equations across dimensions suggests $d=4$ as a sort of critical dimension for the gravitational scattering problem.\footnote{%
  If we may be allowed to speculate, perhaps the change could be  related to the non-existence of CFTs in higher than five spacetime dimensions, in the uplifted holographic picture. 
}

\vskip 0.1in
\noindent
If one managed to solve 
this linearized scattering problem for all even, positive $d$, that would lend credence to the possibility
that all odd $d$ and indeed all real $d>1$ could
be tackled as well.  In the even $d$ case, the fact that the Hankel functions decomposed into plane waves times polynomials meant we could integrate by parts the integral constraint equations (e.g.\ (\ref{firsteqd4}) and (\ref{secondeqd4}) in the $d=4$ case) in order to convert them into ordinary differential equations.  For general $d$, the integral equations do not have a similar Fourier decomposition,
and it naively looks like we would have to solve them directly.  If no clever analytic approach reveals itself, perhaps a numerical solution could be useful. It would also be interesting to explore $d$ analytically 
in various limits, 
for example $d$ near one (the pure $AdS_3$ case) and $d \to \infty$ (see e.g.\ \cite{Betzios:2018kwn,Emparan:2020inr}).  

\vskip 0.1in
\noindent
Further directions include studying multiple thin branes and/or thick interfaces, like in \cite{Baig:2022cnb,Bachas:2022etu}. Multiplying the number of interfaces and/or giving them a thickness may allow for resonant behavior.  
Related boundary-CFT examples, including the $c=1$ boundary sine-Gordon theory \cite{Callan:1993mw}, are known to describe critical dissipative quantum mechanics and barriers in quantum wires. Understanding whether the diffuse angular and evanescent surface channels found here admit a similarly sharp field-theory scattering interpretation would be an interesting direction.

\vskip 0.1in
\noindent
To conclude, the main lesson is that once the pure $AdS_3$  boundary-graviton problem is deformed into a Chamblin-Reall dilaton-gravity problem, interface scattering becomes a genuinely bulk process. The brane's partition of stress-tensor flux depends on more than a single number; 
there is a redistribution into angular and evanescent sectors. In $d=2$ this redistribution is controlled, while in $d=4$ the full story likely requires a black-brane horizon which through introducing a temperature cuts off the infrared in the dual field theory.

\section*{Acknowledgments}

Dedicated to the memory of Andrew Chamblin, Steven Gubser, and Umut G\"{u}rsoy, whose work helped shape the holographic study of nonconformal gravitational backgrounds.
We would like to thank Dionysios Anninos, Costas Bachas, Shira Chapman, Mark Mezei, Shinji Mukohyama, Giuseppe Policastro, and Andy Stergiou for discussion.
C.H. thanks the Oxford Maths Department for hospitality.  
The authors would like to thank the INI for Mathematical Sciences, Cambridge, for support and hospitality during the programme ``BIDS in QFT'', where work on this paper commenced. This work was supported in part by 
STFC grant ST/X000753/1  and EPSRC grant EP/Z000580/1.

\appendix

\section{Effective action}
\label{sec:EA}

It is possible to construct an effective action that is a function just of $\theta_L$ (and where $\theta_R$ is fixed automatically 
in terms of $\theta_L$ by continuity of the metric and dilaton).  
This effective action is not meant to replace the physical variational principle used to derive the junction conditions. It is an auxiliary on-shell functional of the brane angle, designed to reproduce the static interface condition and diagnose stability within the family of scale-covariant embeddings
 The point of this construction is to have a simple way where we can evaluate
\[
\frac{\rd^2 S}{\rd \theta_L^2} \ ,
\]
 and thus whether the solution is a local minimum or maximum in parameter space.
However, to find such an effective action, the variational problem needs to be modified.  Previously, we set up the variational problem in the canonical way where we varied only $g_{ab}$ and $\phi$.  Indeed, the variations of $g_{ab}$ and $\phi$ will depend on $\theta_L$ and so by the chain rule $\delta g_{ab} = (\partial_\theta g_{ab}) \delta \theta$ and $\delta \phi = (\partial_\theta \phi) \delta \theta$, we can expect the 
equations of motion to kill some portions of the total derivative $\frac{ \rd S}{\rd \theta}$.  However,
there are new direct terms in this derivative that did not appear in the original variational problem that come from shifting the range of integration in the classical action.  We expect to pick up a factor of the Lagrangian density  as well as needing to adjust $\delta g_{ab}$ and $\delta \phi$
by their Lie derivatives in the normal direction to account for the change in position of the interface.  These new terms will generically not vanish and thus need to be removed to construct an effective action

\vskip 0.1in
\noindent
These extra terms can be canceled by choosing an auxiliary boundary functional with adjusted coefficients.
We search for an effective action of the form
\begin{equation}
S = S_{\rm bulk} + \frac{1}{2\kappa^2} \int_I \rd^2 x \sqrt{-h} \left(c_K (K_R -K_L) + c_\mu \mu e^{\alpha \phi} \right) \ .
\end{equation}
The choice $c_K = 2$ and $c_\mu = 1$ that we used for the old variational problem does not work.  However, if we choose
\begin{equation}
c_K = d+1 \ , \; \; \; c_\mu = d \ , 
\end{equation}
then we get an effective action whose variation $\frac{\rd S}{\rd \theta_L}  = 0$ is zero when $\theta_L$ and the tension $\mu$ are related in the appropriate way (\ref{alphamurels}).    Moreover, we find that at the critical point
\begin{equation}
\frac{\rd^2 S}{\rd \theta_L^2} = \frac{d^2 (d+1) \ell_B^d \tan^2 \theta_L}{R^{d+1} \kappa^2} ( \cot \theta_L + \cot \theta_R) \ .
\end{equation}
(In evaluating $S$, we are focusing on the $\theta_L$ dependent pieces and ignoring formally infinite but $\theta_L$ independent contributions from the $\rho=0$ boundary that need further regularization.)
Thus the action is convex when $\cot \theta_R + \cot \theta_L > 0$.  Paired with positivity of the tension $\tan \theta_L + \tan \theta_R > 0$,
stability requires the angles to lie in the range $0 \leq \theta_L,  \theta_R < \frac{\pi}{2}$.  In other words, the opening angle between the boundary and the interface must be between 90$^\circ$ and 180$^\circ$ on both sides.  
(A similar result was obtained in the pure $AdS_3$ case in  ref.\ \cite{Czech:2016nxc}.  Note also that for $AdS_3$, $d=1$ and the constants $c_K$ and $c_\mu$ take their canonical values.)

\vskip 0.1in
\noindent
Some identities that underlie this choice of $c_K$ and $c_\mu$ are as follows (focusing on just one side of the interface):
\begin{eqnarray}
(d-1) (K^{\mu\nu} - h^{\mu\nu} K) K_{\mu\nu} &=& - d (\partial^\theta \phi)(\partial_\theta \phi) \ , \\
\partial_\theta  \left( \sqrt{-h}K \right) + \sqrt{-g}(K^{\mu\nu} - h^{\mu\nu} K) K_{\mu\nu} &=& 
\frac{2 d \ell^d}{R^{d+1} \cos^{d+2} \theta} = - \frac{2\kappa^2}{d+1} {\mathcal L}_{\rm bulk} \ ,
\end{eqnarray}
where in the last equality, we introduced the bulk Lagrangian density ${\mathcal L}_{\rm bulk}$.  
Taking $\partial_\theta$ of the Gibbons-Hawking term, like in the second equality above, produces two terms, one of which can
be canceled against $\partial_\theta$ of the bulk action, which is just the bulk Lagrangian ${\mathcal L}_{\rm bulk}$,
 and the other can be converted into an expression in terms of the dilaton $\phi$
and canceled against the tension term in the action, using the on-shell value of $\theta_L$.

\section{Linearized geometric quantities on the interface}
\label{sec:linearization}
We provide a linearized formulation of the gluing problem, following the discussions in \cite{Mukohyama:2000ga}. A different formulation appeared in \cite{Mars:2005ca}, where the geometric quantities were perturbed to second order.
This appendix fixes the geometric conventions used to expand the induced metric and extrinsic curvature on the perturbed CR brane. 
Collecting the results  \cite{Mukohyama:2000ga} here makes the sign and embedding conventions in section \ref{sec:IJ} explicit.

\noindent
\vskip 0.1in
Before perturbing the background, we have an embedding surface $x^\mu(\zeta^a)$ indicating the location of the interface, an embedding frame $e^\mu_a$, an induced metric $h_{ab}$ and an extrinsic curvature $K_{ab}$ 
\be
e^{\mu}_{a} = \frac{\der x^\mu}{\der \zeta^a}\,,~~~ h_{ab} = e^{\mu}_a e^\nu_b g_{\mu\nu}\,, ~~~ K_{ab} =e^\mu_a e^\nu_b \nabla_{\mu}n_\nu = \frac{1}{2} e^\mu_a e^\nu_b \mL_{n} g_{\mu\nu}\,.
\ee
Given a general metric fluctuation $g_{\mu\nu} \to g_{\mu\nu} + \delta g_{\mu\nu}$, the interface has to be deformed $x^\mu (\zeta^a ) \to x^\mu (\zeta^a ) + \xi^\mu (\zeta^a) $ to allow for a possible gluing. The surface deformation parameter $\xi^\mu$ and the metric fluctuation $\delta g_{\mu\nu}$ are treated as the same order of perturbation. At linearized order, those quantities are found to be 
\begin{align}
\delta e^\mu_a &{= -\mL_\xi e^{\mu}_a = -(\xi^{\nu} \nabla_\nu e^{\mu}_a - e^\nu_a \nabla_\nu \xi^\mu})\,, \\
\delta h_{ab} & =\xi^\alpha \der_\alpha ( e^{\mu}_a e^\nu_b  g_{\mu\nu}) + \delta(e^\mu_a e^\nu_b) g_{\mu\nu} +
 e^{\mu}_a e^\nu_b \delta g_{\mu\nu} =  e^{\mu}_a e^\nu_b(\mL_{\xi} g_{\mu\nu} + \delta g_{\mu\nu})\,,\\
 \label{eq:deltaKab}
\delta K_{ab} 
&= \frac{1}{2} e^{\mu}_a e^\nu_b \left( \mL_{\xi} \mL_n g_{\mu\nu} +\nabla_\mu \delta n_\nu + \nabla_\nu \delta n_\mu  - 2n^\alpha \delta\Gamma_{\alpha \mu \nu} \right) \,,
\end{align}
with the  perturbed quantities
\begin{align}
\delta n_\mu &= \frac{1}{2}n_\mu n^\nu n^\gamma \delta g_{\nu \gamma} - h_{\mu\nu} e^{\nu}_a h^{ab}n_\gamma \delta e^\gamma_b\,,~~~ h^{\mu\nu} = e^\mu_a h^{ab} e^\nu_b\,,\\
\delta\Gamma_{\alpha \mu \nu} &= \frac{1}{2}\left( \delta g_{\alpha \mu;\nu}  +   \delta g_{\alpha \nu;\mu} -  \delta g_{ \mu\nu; \alpha} \right)\,.
\end{align}
Substituting the explicit form of $\delta e^\mu_a$, the variation of the normal vector can be further expressed as
\begin{align}
\delta n_\mu &= \frac{1}{2}n_\mu n^\nu n^\gamma \delta g_{\nu \gamma}-(g_{\mu\nu} - n_\mu n_\nu)(\xi^\alpha \nabla_\alpha n^\nu + n_\alpha \nabla^{\nu} \xi^\alpha)\no\\
&=-\mL_\xi n_\mu + n_\mu n^\alpha n^\beta \left(\nabla_\alpha \xi_\beta + \frac{1}{2}\delta g_{\alpha\beta} \right)\,,
\end{align}
while the double Lie derivative term in \eqref{eq:deltaKab} can be expanded using the commutating relation between the Lie and covariant derivatives \cite{yano1957theory}
\begin{align}
 [\mL_\xi, \nabla_\mu] n_\nu =-n_\alpha\nabla_\mu \nabla_\nu \xi^{\alpha} + n_\alpha { R_{\lambda \mu\nu}}^\alpha \xi^\lambda\,,
\end{align}
giving an explicit form of the variation of the extrinsic curvature \cite{Mukohyama:2000ga,Mars:2005ca}, 
\begin{align}
\delta K_{ab} = & \frac{1}{2} n^{ \mu} n^{ \nu}\left(\delta g_{\mu\nu}+2\xi_{\mu ; \nu}\right) K_{ab}\no \\
& -\frac{1}{2} n^{ \lambda} e_a^{\mu} e_b^{\nu}\left[2 \delta \Gamma_{\lambda \mu \nu}+\xi_{\lambda ; \mu\nu }+\xi_{\lambda ; \nu\mu}+\left(R_{\alpha \mu\lambda \nu}+R_{\alpha \nu \lambda \mu}\right) \xi^{\alpha}\right]\,.
\end{align}

\section{Embedding function solutions}
\label{sec:solns}

The main text determines the angular amplitudes $\tilde R$ and $\tilde T$.
 Here we record the corresponding embedding functions and gauge-mode choices. These expressions are not needed for the flux discussion but verify that the full linearized junction problem can be solved with continuous embedding data at $\rho = 0$.

\subsection{Solutions for $d=2$}

\vskip 0.1in
\noindent
We have not provided the actual solutions for the embedding functions nor discussed the gauge solutions.  Regarding the gauge solutions,
if we choose the $k_G$ wave vectors to be the same as the physical ones,\footnote{%
 There appears to be some arbitrariness in our solution.  The resolution we describe above is the shortest and simplest. Another strategy would be to tune the wave vector of the gauge solution such that when projected onto the interface, it has the same Fourier component as the physical fluctuations.  However, it is not possible to make the embedding functions continuous at the $\rho=0$ boundary with this alternate choice.  That said, if we also include the gauge solutions corresponding to the $c_2$ integration constant, we can ensure continuity, but the solutions begin to look very messy.
} we can make $\zeta$, $\lambda$, and $\delta$ continuous at $\sigma=0$ if we set
\begin{equation}
I_G =   \frac{i}{\omega^3 \cos \alpha_I}I \ , \; \; \;
R_G =   \frac{i} {\omega^3 \cos \alpha_R}R \ , \; \; \;
T_G =   \frac{i}{\omega^3  \cos \alpha_T} T \ .  
\end{equation}
With these choices $\Delta \sim O(\sigma^2)$, $\delta \sim O(\sigma)$, $\lambda \sim O(\sigma^2)$, $\zeta \sim O(\sigma)$. 
Note $\Delta$ already vanishes to $O(\sigma^2)$ without tuning the gauge solution.  

\vskip 0.1in
\noindent
The final solutions for the embedding functions are
\begin{eqnarray}
\frac{\rd \lambda}{\rd \myangle} &=& \frac{ e^{-i \omega \sigma \sin \myangle}}{\ell_B \omega^2}
\biggl(  \left( -i \sigma \omega - \left(1-e^{i \sigma \omega \cos \theta_L \sin (\theta_L+\myangle)} \right) \csc(\theta_L+\myangle) \sec \theta_L \right)\tilde I
\nonumber \\
&&
+  \left( -i \sigma \omega + \left(1- e^{-i \sigma \omega \cos \theta_L \sin (\theta_L-\myangle)} \right) \csc(\theta_L-\myangle) \sec \theta_L \right) \tilde R
\nonumber \\
&&
+  \left( i \sigma \omega - \left(1- e^{-i \sigma \omega \cos \theta_R \sin (\theta_R-\myangle)} \right) \csc(\theta_R-\myangle) \sec \theta_R \right) \tilde T
 \biggr) \ .
\\
\frac{\rd \zeta}{\rd \myangle} &=& \frac{e^{-i \omega \sigma \sin \myangle}}{\ell_B \omega^2} \biggl(
\left( -i \sigma \omega \sin \myangle + \left(1 - e^{i \sigma \omega \cos \theta_L \sin (\theta_L+\myangle)} \right)\cot(\theta_L+\myangle) \tan \theta_L \right)
\tilde I
\nonumber \\
&& \left( - i \sigma \omega \sin \myangle + \left(1- e^{-i \sigma \omega \cos \theta_L \sin (\theta_L-\myangle)} \right) \cot(\theta_L-\myangle) \tan \theta_L \right) \tilde R
\nonumber \\
&&
 \left( i \sigma \omega \sin \myangle - \left(1- e^{-i \sigma \omega \cos \theta_R \sin (\theta_R-\myangle)} \right) \cot(\theta_R-\myangle) \tan \theta_R \right) \tilde T
 \biggr) \ , \\
 \frac{\rd \delta}{\rd\myangle} &=& \frac{ e^{-i \omega \sigma \sin \myangle}}{\ell_B \omega^2} \biggl(
 \left(1-e^{i \sigma \omega \cos \theta_L \sin (\theta_L+\myangle)} \right) \cot(\theta_L+\myangle) \tilde I  \nonumber \\
&&+ \left(1- e^{-i \sigma \omega \cos \theta_L \sin (\theta_L-\myangle)} \right)  \cot (\theta_L-\myangle) \tilde R \nonumber \\
&&+ \left(1- e^{-i \sigma \omega \cos \theta_R \sin (\theta_R-\myangle)} \right)  \cot (\theta_R-\myangle) \tilde T
 \biggr) \ , \\
\frac{\rd \Delta}{\rd \myangle} &=& \frac{1}{2 \ell_B} \sigma^2 e^{-i \omega \sigma \sin \myangle} (- \tilde I - \tilde R + \tilde T) \ .
\end{eqnarray}
These functions all have engineering dimensions of length, as anticipated.
Note the $e^{-i \sigma \omega \cos \theta \sin(\theta - \myangle)}$ dependence in the solution above does not have nice growth 
on the evanescent branch as $\sigma \to \infty$.  As this plane wave factor comes from the gauge modes, the growth appears to be a gauge artifact.

\subsection{Solutions for $d=4$}

\vskip 0.1in
\noindent
The actual solution in this case takes the form
\begin{eqnarray}
\frac{\rd \lambda}{\rd \myangle} &=& \frac{ e^{-i \sigma \omega \sin \myangle}}{\gamma \omega^3 \ell_B^2} \biggl(
\left( -\omega^2 \sigma^2 + 3 i \sigma \omega \csc(\theta_L+\myangle)\sec \theta_L +3 \left(1-e^{i \sigma \omega \cos \theta_L \sin(\theta_L+\myangle)}\right)
\csc(\theta_L+\myangle)^2 \sec^2 \theta_L \right) \tilde I \nonumber \\
&&
+ \left( -\omega^2 \sigma^2  - 3 i \sigma \omega \csc(\theta_L-\myangle)\sec \theta_L +3 \left(1-e^{-i \sigma \omega \cos \theta_L \sin(\theta_L-\myangle)}\right)
\csc(\theta_L-\myangle)^2 \sec^2 \theta_L \right) \tilde R \nonumber \\
&&
+ \left( \omega^2 \sigma^2 + 3 i \sigma \omega \csc(\theta_R-\myangle)\sec \theta_R -3 \left(1-e^{-i \sigma \omega \cos \theta_R \sin(\theta_R-\myangle)}\right)
\csc(\theta_R-\myangle)^2 \sec^2 \theta_R \right) \tilde T
\biggr) \ , \\
\frac{\rd \zeta}{\rd \myangle} &=& \frac{ e^{-i \sigma \omega \sin \myangle}}{ \gamma \omega^3 \ell_B^2} \biggl(
\bigl( i \sigma \omega(1 + i \sigma \omega \sin \myangle) - 3 i \sigma \omega \cot (\theta_L + \myangle) \tan \theta_L  \nonumber \\
&& \hskip 0.75in - 3\left(1 - e^{i \sigma \omega \cos \theta_L \sin(\theta_L+\myangle)}\right) \cot(\theta_L + \myangle) \csc (\theta_L + \myangle) \sec \theta_L \tan \theta_L \bigr) \tilde I \nonumber \\
&&
 \hskip 0.5in  + \bigl( i \sigma \omega (1 + i \sigma \omega \sin \myangle) - 3 i \sigma \omega \cot (\theta_L - \myangle) \tan \theta_L   \nonumber \\ 
&&  \hskip 0.75in+ 3\left(1 - e^{-i \sigma \omega \cos \theta_L \sin(\theta_L-\myangle)}\right) \cot(\theta_L - \myangle) \csc (\theta_L - \myangle)  \sec \theta_L \tan \theta_L \bigr) \tilde R\nonumber \\
&&
 \hskip 0.5in + \bigl( -i \sigma \omega  (1 + i \sigma \omega \sin \myangle) + 3 i \sigma \omega \cot (\theta_R - \myangle) \tan \theta_R   \nonumber \\
&&
  \hskip 0.75in - 3\left(1 - e^{-i \sigma \omega \cos \theta_R \sin(\theta_R-\myangle)}\right) \cot(\theta_R - \myangle) \csc (\theta_R- \myangle) \sec \theta_R \tan \theta_R \bigr) \tilde T
\biggr) \ , \\
\frac{\rd \delta}{\rd \myangle} &=&  \frac{ e^{-i \sigma \omega \sin \myangle}}{\gamma \omega^3 \ell_B^2} \biggl(
3 \cot (\theta_L + \myangle) \left(  - i \sigma \omega  - \left(1 - e^{i \sigma \omega \cos \theta_L \sin(\theta_L+\myangle)}\right) \csc (\theta_L+\myangle)\sec \theta_L \right) \tilde I \nonumber \\
&&
+3 \cot (\theta_L - \myangle) \left(  - i \sigma \omega + \left(1 - e^{-i \sigma \omega \cos \theta_L \sin(\theta_L-\myangle)}\right) \csc (\theta_L-\myangle) \sec \theta_L \right) \tilde R \nonumber \\
&&
+3 \cot (\theta_R - \myangle) \left(  - i \sigma \omega + \left(1 - e^{-i \sigma \omega \cos \theta_R \sin(\theta_R-\myangle)}\right) \csc (\theta_R-\myangle) \sec \theta_R \right)  \tilde T
\biggr) \ , \\
\frac{\rd \Delta}{\rd \myangle} &=& \frac{e^{-i \sigma \omega \sin \myangle} }{4 \gamma  \ell_B^2 } i\sigma^3 \left( \tilde I \cos \theta_L  + \tilde R \cos \theta_L 
-\tilde T \cos \theta_R  \right) \ .
\end{eqnarray}
Note all these functions have engineering dimensions of length, as required.
The gauge solution has been chosen such that
\begin{equation}
I_G = i \frac{2 \gamma  }{ \omega^4 \cos^2 \alpha_I} I \ , \; \; \;
R_G =  i \frac{2 \gamma  }{ \omega^4 \cos^2 \alpha_R} R \ , \; \; \;
T_G = i  \frac{2 \gamma }{ \omega^4 \cos^2 \alpha_T} T \ ,
\end{equation}
and the embedding functions are continuous at $\rho = 0$.  
Near $\rho=0$, these functions behave as $\lambda(\rho) = O(\rho^2)$, $\zeta(\rho) = O(\rho)$, $\delta(\rho) = O(\rho^2)$, and $\Delta(\rho) = O(\rho^3)$.

\section{Roots of a Cubic}
\label{sec:cubic}

The roots listed here are those of the characteristic cubic governing the $d=4$ angular response functions in section \ref{sec:d4}.
The solutions to the cubic equation (\ref{mycubic}) can be written explicitly:
\begin{align}
\kappa_1 = \frac{2}{3}m-2\lambda\,,~~~ \kappa_2 = \frac{2}{3}m+\lambda + i \sqrt{1+\frac{m^2}{6}+3\lambda^2}\,,~~~\kappa_3 = \frac{2}{3}m+\lambda - i \sqrt{1+\frac{m^2}{6}+3\lambda^2}\,,
\end{align}
where $\lambda\in \mathbb{R}$ is the real solution of the qubic equation
\be
\lambda ^3+\frac{1}{24} \lambda  \left(m^2+6\right)-\frac{m D}{1728}=0\,,~~~D=7 m^2+81 p^2+36\,,
\ee
of which the real one is given as 
\be
\lambda = \frac{\sqrt[3]{\Delta }}{24}-\frac{m^2+6}{3 \sqrt[3]{\Delta }}\,,~~~\Delta=4 \left(\sqrt{D^2 m^2+32 \left(m^2+6\right)^3}+D m\right)\,.
\ee

\bibliographystyle{JHEP}
\bibliography{bib}

\end{document}